\documentclass[a4paper,11pt]{article}
\usepackage{amsmath}
\usepackage{amsfonts}
\usepackage{amsthm}
\usepackage{amssymb}
\usepackage{amscd}
\usepackage[british]{babel}
\usepackage{graphicx}
\usepackage{psfrag}
\usepackage{epsfig}
\usepackage{rotating}
\usepackage{times}

\theoremstyle{plain}

\newtheorem{remark}{Remark}[section]

\newcommand{\boxend}{\flushright{$\Box$}}

\renewcommand{\tilde}{\widetilde}

\usepackage{color}

\begin{document}

\title{Bounce Loop Quantum Cosmology Corrected Gauss-Bonnet Gravity}
\author{
J. Haro$^{1}$\,\thanks{jaime.haro@upc.edu}
A.N. Makarenko$^{2, 3}$\,\thanks{andre@tspu.edu.ru}
A.N. Myagky$^{4}$\,\thanks{alex7604@mail.ru}
S. D. Odintsov$^{5,6}$\,\thanks{odintsov@ieec.uab.es}
V.K. Oikonomou$^{3,7}$\,\thanks{v.k.oikonomou1979@gmail.com} \\ \\
$^{1}$  Universitat Polit\`ecnica de Catalunya,\\  08028 Barcelona,  , Spain\\
$^{2}$ National Research Tomsk State University,\\ 634050 Tomsk, Russia\\
$^{3}$ Tomsk State Pedagogical University, 634061 Tomsk, Russia\\
$^{4}$ National Research Tomsk Polytechnic University, 634050 Tomsk, Russia\\
$^{5}$ Institut de Ciencies de lEspai (IEEC-CSIC),\\
Campus UAB, Carrer de Can Magrans, s/n\\
08193 Cerdanyola del Valles, Barcelona, Spain\\
$^{6}$ ICREA,\\
Passeig Llu{\'\i}­s Companys, 23,
08010 Barcelona, Spain\\
$^{7}$ Lab. Theor. Cosmology, \\
Tomsk State University of Control Systems
and Radioelectronics,\\ 634050 Tomsk, Russia (TUSUR)}

\maketitle

{%$^{a,c}$Instituto de Ciencias del Espacio (CSIC) and
%Institut d'Estudis Espacials de Catalunya (IEEC/CSIC)\\ Universitat
%Aut\`{o}noma de Barcelona, Torre C5-Parell-2a planta, 08193 Bellaterra
%(Barcelona) Spain\\

%$^b$Departament de Matem\`atica Aplicada I, Universitat
%Polit\`ecnica de Catalunya, Diagonal 647, 08028 Barcelona, Spain \\
}

\thispagestyle{empty}

\begin{abstract}
We develop an effective Gauss-Bonnet extension of Loop Quantum Cosmology, by introducing holonomy
corrections in modified $F(\mathcal{G})$ theories of gravity. Within the context of our formalism, we provide a perturbative expansion in the critical density,
a parameter characteristic of Loop Quantum Gravity theories, and we result in having leading order corrections to the classical $F(\mathcal{G})$ theories of gravity. After
extensively discussing the formalism, we present a reconstruction method that makes possible to find the Loop Quantum Cosmology corrected $F(\mathcal{G})$ theory that can
realize various cosmological scenarios. We exemplify our theoretical constructions by using bouncing cosmologies, and we investigate which Loop Quantum Cosmology corrected
Gauss-Bonnet modified gravities can successfully realize such cosmologies.
\end{abstract}

\vspace{0.5cm}

{\bf Pacs numbers:} 04.50.Kd; 04.60.Pp; 40.60.Ds

%\vspace{0.5cm}

\section{Introduction}

The Big Bang era is one of the less understood periods of the evolution of our Universe, and the physics behind this era is still inconceivable. The standard approach that provides a classical physical description is based on the assumption that our Universe is described by a Friedman-Lemaitre-Robertson-Walker (FLRW) metric, which captures most of the present date characteristics of our Universe, i.e. its space uniformity, large scale homogeneity etc. However, the classical cosmological approach leads inevitably to an initial singularity, which is a rather ``embarrassing'' feature of the classical description, because due to this singularity, the closed time-like geodesics which pass from this singularity, have a finite proper length, but no end points to normal space away from the singularity. In addition to this, there has been conjectured for some years \cite{penrose}, that naked singularities should be well hidden behind horizons, the so-called cosmic censorship conjecture. However, no formal proof
exists up to date for this cosmic censorship hypothesis, nevertheless it is considered a physically appealing hypothesis. From a cosmological initial singularity exists, then it belongs to a period of time that quantum physics, or more appropriately, quantum cosmology governs the physical phenomena.

One of the most appealing quantum cosmological theories is  holonomy corrected Loop Quantum Cosmology (LQC) \cite{LQC}, which is promising from various points of view. For an important stream of papers and reviews on this fastly developing research topic, consult \cite{LQC}, and references therein. In the context of LQC, many theoretical inconsistencies or ellipses of the classical cosmological theories, could consistently be explained. Having such a promising theoretical framework at hand, it is compelling to investigate theoretical modifications, along the research lines of modified gravity. In this context, $F(R)$ extensions of LQC have recently been developed in \cite{zm11,zm11a}, where, following the idea of \cite{Risi},
holonomy corrections are introduced in Einstein frame (EF), because in that frame the gravitational
part of the Hamiltonian is linear in the scalar curvature and thus the procedure to introduce these effects is the same as in General Relativity.
These corrections have been applied to $R^2$ gravity  \cite{aho}, obtaining a bouncing model free of singularities. On the other hand,  inspired in the introduction
of holomy effects in the theory of cosmological perturbations \cite{bojowald}, an effective  way to introduce these corrections in $F(R)$ gravity was developed in \cite{haro14}, where the EF formulation is not necessary to be used. The main characteristic of it is that it could be applied to other theories, where there is not a conformal equivalence with EF, and therefore allowing to introduce these corrections to Gauss-Bonnet gravity in an explicit way. Gauss-Bonnet gravity is characterized by the Gauss-Bonnet invariant curvature which we denote ${\mathcal G}$ in this paper.

Adopting the research lines of \cite{haro14}, we perform a Gauss-Bonnet an effective extension of LQC. Once we have obtained this new theory, we generalize the reconstruction method obtained in \cite{bmmo}, and we reconstruct some  holonomy corrected LQC-$F({\mathcal G})$
gravity theories. We shall focus our reconstruction procedure to realizing mainly bouncing cosmologies. The reason for that specialization is that bouncing cosmologies have the very appealing feature of the absence of an initial singularity \cite{bounces,bounce1,bounce2,bounce3,bounce4,bounce5,quintombounce,matterbounce,superbounce2,superbounce3}, plus accelerating expansion can consistently be described in the context of these theories \cite{ekpyr1,ekpyr2,ekpyr3}. These kind of cosmologies are very interesting owing to the fact that nowadays they represent the most promising alternative theory to the inflationary paradigm, because if at very early times the Universe is nearly matter dominated in the contracting phase, the power-spectrum of the modes that leave the Hubble radius during this regime will be almost flat \cite{powerspectrum}. Moreover, is has been shown in \cite{eho} that bouncing cosmologies provide theoretical values of the cosmological parameters, such as the spectral index and its running,
that fit well with recent Planck data \cite{planck}.

But what is the motivation to introduce holonomy corrections in an $F(\mathcal{G})$ gravity in the first place. We need therefore to discuss the motivation to include LQC effects in $F(\mathcal{G})$, since Gauss-Bonnet as a kind of effective limit of string theory, and the Einstein-Hilbert LQC corrected gravity is finite, so why introducing LQC effected $F(\mathcal{G})$ gravity. The answer is that, since the non-LQC $F(\mathcal{G})$-gravity contains finite time singularities, which in some cases can be crushing type singularities. So a necessary extension of standard $F(\mathcal{G})$ gravity to include LQC effects is necessary to see if any new effects that the LQC corrections might bring along. For example, recently in Ref. \cite{oikofg}, it has been pointed out that for a Type IV singular bounce, near the bounce point, in which case all the modes that are relevant for today observations are well inside the Hubble radius, the power spectrum of primordial curvature perturbations is not scale invariant. Thereby, a quite important task is to investigate whether the spectrum of primordial curvature perturbations becomes scale invariant by including LQC holonomy corrections in the classical $F(\mathcal{G})$ gravity. The same case applies for the Einstein frame LQC corrected $F(R)$ gravity, with a thorough investigation of this case being performed in \cite{aho}. Hence, the LQC corrected $F(\mathcal{G})$ is necessary for the aforementioned reasons, and also in order to see whether the singularity structure of the cosmological dynamical system is changed at some extent.

The paper is organized as follows: In section 2, using the method of Lagrange multipliers, we construct the classical dynamical equations in Gauss-Bonnet gravity, for a flat FLRW geometry. In section 3, we review the way to introduce holonomy corrections
in standard Einstein-Hilbert general relativity, for the particular case of the flat FLRW geometry. In section 4, we extend  LQC to Gauss-Bonnet gravity obtaining a holonomy corrected Friedmann equation, which contains all the dynamical information of the system.
Section 5 is devoted to explain the reconstruction method in holonomy corrected Gauss-Bonnet gravity. Sections 6 and 7 contain examples of reconstruction, for power law, bouncing exponential models and generic bouncing models respectively. The conclusions follow at the end of the paper.

\section{ Gauss-Bonnet $F({\mathcal G})$ gravity ''a la Ostrogradsky''}

When one considers the flat FLRW geometry, the Lagrangian of Gauss-Bonnet
 $F({\mathcal G})$ gravity in the vacuum is given by (in units $\hbar=c=8\pi G=1$),
\begin{eqnarray}\label{A1}
 {\mathcal L}(V,\dot{V},\ddot{V})=\frac{1}{2}V (R+F({\mathcal G})),
\end{eqnarray}
where $V=a^3$ is the volume and $\mathcal{G}$ denotes the Gauss-Bonnet curvature, which can be expressed in terms of the volume and its higher derivatives,
\begin{equation}\label{gbcurv}
{\mathcal G}=24H^2(\dot{H}+H^2)=\frac{8\dot{V}^2}{3V^2}\left(\frac{\ddot{V}}{3V}-\frac{2\dot{V}^2}{9V^2}\right)
\end{equation}
Ostrogradsky's idea to obtain the Hamiltonian from a Lagrangian containing higher order derivatives, is epitomized in the introduction of a Lagrange multiplier term in the Lagrangian, which we denote $\mu$, in the following way \cite{dsy,haro14}:
\begin{eqnarray}\label{A2}
 {\mathcal L}_1(V,\dot{V},\ddot{V},{\mathcal G})=\frac{1}{2}V (R+F({\mathcal G}))+
 \mu \left(\frac{8\dot{V}^2}{3V^2}\left(\frac{\ddot{V}}{3V}-\frac{2\dot{V}^2}{9V^2}\right)-{\mathcal G}\right).
\end{eqnarray}
Maximizing with respect to the Gauss-Bonnet invariant ${\mathcal G} $ yields $\mu=\frac{1}{2}VF'({\mathcal G})$. In order for the second derivative of $V$ to be removed, we subtract from the Lagrangian the following
total derivative term,
\begin{equation}\label{totalderiv}
\frac{d}{dt}\left(\frac{4\dot{V}^3}{27V^2}F'({\mathcal G})+\dot{V}\right)\, ,
\end{equation}
which does not change the dynamics of the system, and by replacing the Lagrange multiplier $\mu$ by its value $\frac{1}{2}VF'({\mathcal G})$, we finally obtain,
\begin{eqnarray}\label{A3}
 \tilde{\mathcal L}(V,\dot{V},{\mathcal G},\dot{{\mathcal G}})=-\frac{\dot{V}^2}{3V}+\frac{1}{2}VF({\mathcal G})
 %\nonumber\\
 -\frac{1}{2}VF'({\mathcal G}){\mathcal G}- \frac{4\dot{V}^3}{27V^2}F''({\mathcal G})\dot{\mathcal G}\, .
\end{eqnarray}
As it is obvious by Eq. (\ref{A3}), we can see that the Lagrangian (\ref{A3}) depends on the variables $(V,{\mathcal G})$ and their first derivatives. We can therefore obtain the corresponding canonically conjugate momenta, which are,
\begin{eqnarray}\label{A4}
 p_V\equiv \frac{\partial\tilde{\mathcal L}(V,\dot{V},{\mathcal G},\dot{{\mathcal G}})}{\partial \dot{V}}=-\frac{2\dot{V}}{3V}-\frac{4\dot{V}^2}{9V^2}F''({\mathcal G})\dot{\mathcal G},
 \nonumber\\
 p_{\mathcal G}\equiv \frac{\partial\tilde{\mathcal L}(V,\dot{V},{\mathcal G},\dot{{\mathcal G}})}{\partial \dot{{\mathcal G}}}=-\frac{4\dot{V}^3}{27V^2}F''({\mathcal G}),
\end{eqnarray}
and consequently the classical gravitational part of the Hamiltonian becomes,
\begin{eqnarray}\label{A5}
 {\mathcal H}_{grav}(V,{\mathcal G},p_V,p_{\mathcal G})\equiv \dot{V}p_V+\dot{{\mathcal G}}p_{\mathcal G}-\tilde{\mathcal L}(V,\dot{V},{\mathcal G},\dot{{\mathcal G}})\nonumber\\
 =\frac{3}{V}\left(\frac{V^2p_{\mathcal G}}{4F''({\mathcal G})} \right)^{2/3}-3p_V\left(\frac{V^2p_{\mathcal G}}{4F''({\mathcal G})} \right)^{1/3}+\frac{V}{2}\left({\mathcal G}F'({\mathcal G})-F({\mathcal G})\right).
\end{eqnarray}
The Hamiltonian constraint ${\mathcal H}_{grav}(V,{\mathcal G},p_V,p_{\mathcal G})=0$ leads to the well-known modified Friedmann equation in the vacuum for non-LQC corrected $F({\mathcal G})$ gravity, which is,
\cite{nojiri}
\begin{eqnarray}\label{A6}
 6H^2+24H^3\dot{{\mathcal G}}F''({\mathcal G})-{\mathcal G}F'({\mathcal G})+F({\mathcal G})=0\, .
\end{eqnarray}
 Note that in ordinary Einstein-Hilbert gravity, where $F({\mathcal G})=0$, the canonically conjugate momenta are equal to,
 \begin{equation}\label{canmomeinstei}
 p_V=-\frac{2\dot{V}}{3V}=-2H,\, \, \,\frac{p_{\mathcal G}}{F''({\mathcal G})}=\frac{1}{2}p_V^3V
\end{equation}
and therefore, the classical gravitational part of the Hamiltonian takes the following simplified form,
 \begin{eqnarray}\label{A7}
 {\mathcal H}_{grav}(V,p_V)=-\frac{3}{4}p_V^2V=-3H^2V\, .
 \end{eqnarray}

 \begin{remark}
Note that, the Hamilton equations $\dot{V}=\frac{\partial{\mathcal H}_{grav}}{\partial p_V} $ and $\dot{\mathcal G}=\frac{\partial{\mathcal H}_{grav}}{\partial p_{\mathcal G}} $
are simple identities. The equation $\dot{p_{\mathcal G}}=-\frac{\partial{\mathcal H}_{grav}}{\partial {\mathcal G}} $ is equivalent to the modified Friedmann equation (\ref{A6}), and
the equation $\dot{p_{ V}}=-\frac{\partial{\mathcal H}_{grav}}{\partial V} $ which corresponds to the modified Raychauduri equation in $F({\mathcal G})$ gravity, could be obtained
taking the derivative with respect to the cosmic time of the modified Friedmann equation. As a consequence, in the vacuum the dynamics in  $F({\mathcal G})$ gravity is only modelled
by the modified Friedmann equation (\ref{A6}), and when one considers matter one has to include the conservation equation $\dot{\rho}=-3H(\rho+P)$, being $P$ the pressure.
 \end{remark}

Before ending this section, we need to stress an important issue: Since in principle there are infinitely many canonical transformations, this means that the modified $F({\mathcal G}) $ gravity could be formulated using infinitely many sets of variables (two coordinates
and their corresponding canonically conjugate momenta). Note that some of these sets of variables will be meaningless physically speaking, because they are built using a combination of both coordinates and momenta, giving new quantities with a very difficult physical interpretation. Moreover, since the introduction of holonomy effects critically depend on the set of variables used, in effect there are
infinitely many ways to introduce
holonomy effects in modified $F({\mathcal G})$ gravity. Consequently, there will be infinitely many different effective holonomy corrected Friedmann equations in $F({\mathcal G})$ gravity.

 \section{Introduction of holonomy corrections}

 In this section and in order for the article to be maintained self-complete, we shall describe the technique of introducing holonomy corrections in standard Einstein-Hilbert gravity. Assuming a flat FLRW geometry, first of all one can consider the variable $\beta=-\frac{\gamma}{2}p_V=\gamma H$
 (\cite{s09}), where $\gamma$ is the Barbero-Immirzi parameter. In terms of the parameter $\beta$, the Hamiltonian (\ref{A7})
becomes ${\mathcal H}_{grav}(V,\beta)=-\frac{3\beta^2}{\gamma^2}V$. However,
in LQC, due to the discrete nature of space, the quantum operator $\hat{\beta}$ is not well defined
(see for instance \cite{h12} or \cite{as11} for a status report on this issue). Then, in order to build the quantum theory, the gravitational part of the Hamiltonian must be redefined. To be precise, we will consider holonomies of the form $
h_j(\lambda)\equiv e^{-i\frac{\lambda \beta}{2}\sigma_j}$, where
$\sigma_j$ denote the Pauli matrices and $\lambda$ is the square root of the minimum eigenvalue of the area operator in loop quantum gravity.
Since $\beta^2$ does not have a well-defined quantum operator, in order for a consistent quantum Hamiltonian operator to be constructed, an almost periodic function that approaches $\beta^2$ for small values of $\beta$ is needed. This can be done using the general formulae of loop quantum gravity to obtain the holonomy corrected Hamiltonian, which is equal to,
\begin{eqnarray}\label{ham}
&& \hspace*{-5mm}
{\mathcal H}_{hol,grav}(V,\beta)\equiv-\frac{2 {V}}{\gamma^3 \lambda^3}
\sum_{i,j,k}\varepsilon^{ijk} Tr\left[
h_i(\lambda)h_j(\lambda)h_i^{-1}(\lambda)
\right. \nonumber \\ && \left. \times
 h_j^{-1}(\lambda)h_k(\lambda)\{h_k^{-1}(\lambda),{V}\}\right]\, .
\end{eqnarray}
The Hamiltonian (\ref{ham}) captures the underlying loop quantum dynamics, and for a detailed account on this issue see for instance \cite{aps06}).

A simple calculation \cite{bo08,he10,dmw09} shows that the Hamiltonian of Eq. (\ref{ham}) acquires the simple form
\begin{eqnarray}\label{A8}
  {\mathcal H}_{hol,grav}(V,\beta)=-3\frac{\sin^2(\lambda \beta)}{\lambda^2\gamma^2}V,
 \end{eqnarray}
which indicates that effectively, holonomy effects can be introduced by explicitly performing the replacement
$\beta\rightarrow \frac{\sin(\lambda \beta)}{\lambda}$ or equivalently $p_V\rightarrow -\frac{2\sin(\lambda \beta)}{\lambda\gamma}$.

In order to obtain the holonomy corrected Friedmann equation, in principle it is necessary to use the full Hamiltonian, i.e., the Hamiltonian that contains both, gravitational and matter parts,
\begin{eqnarray}\label{A9}{\mathcal H}_{hol}(V,\beta) =-3\frac{\sin^2(\lambda \beta)}{\lambda^2\gamma^2}V+\rho V,\end{eqnarray}
Then, by combining the Hamilton equation,
\begin{eqnarray}\label{A10}
 \dot{V}=-\frac{\gamma}{2}\frac{\partial{\mathcal H}_{hol}(V,\beta) }{\partial \beta}=-3V\frac{\sin(2\lambda \beta)}{2\lambda\gamma},
\end{eqnarray}
with the Hamiltonian constraint ${\mathcal H}_{hol}(V,\beta) =0$, we obtain the well-known modified Friedmann equation \cite{Friedmann},
\begin{eqnarray}\label{A11}
 H^2=\frac{\rho}{3}\left(1-\frac{\rho}{\rho_c}\right),
\end{eqnarray}
where $\rho_c\equiv\frac{3}{\lambda^2\gamma^2}$ is the so-called {\it critical density}. Practically this density is what measures the
{\it strength} of the loop quantum effects in cosmology, so if it is infinite, and for finite energy densities, the holonomy corrected Friedmann equation will
be reduced to its well known form
in standard cosmology.

\section{{Holonomy corrected  $F({\mathcal G})$ gravity}}

Recently, holonomy correction are introduced in $F(R)$ gravity using the property that $F(R)$ gravity in the Jordan Frame (JF) is equivalent to General Relativity in the
Einstein Frame (EF). Then, introducing, as in standard LQC, holonomy corrections in EF and comming back to the JF one obtains what is named as Loop Quantum $F(R)$
gravity \cite{zm11,zm11a}.

Unfortunately, Gauss-Bonnet gravity is not equivalent to General Relativity in any frame. Then, to introduce holonomy corrections, other strategy must be used.
What we will set up is an effective LQC theory for $F({\mathcal G})$ gravity based in this key point:
As we have seen in previous
Section,
stardard holohomy corrected LQC, could essentially be obtained from the replacement
$\beta\rightarrow \frac{\sin(\lambda \beta)}{\lambda}$ or equivalently $p_V\rightarrow -\frac{2\sin(\lambda \beta)}{\lambda\gamma}$.
Therefore, when  there  is not an established  way to perform a kinematical loop quantization of  the phase space, as  when one deals with cosmological perturbations,
%Moreover, following this idea,
%since, at that moment, there is not a way to perform a kinematical loop quantization of  the phase space corresponding to cosmological perturbations,
 in order to incorporate loop
corrections,  an effective Hamiltonian is built  adopting as a prescription the replacement of the Ashtekar connection by a suitable function of that connection \cite{bojowald}.
Then, with the same spirit, we will introduce  holonomy corrections in  general $F({\mathcal G})$  using the following  recipe:

To introduce the holonomy correction in  general $F({\mathcal G})$ we will adopt the following recipe:
In analogy with the standard Einstein-Hilbert gravity case, where $F({\mathcal G})=0$, we will replace the momentum that in Einstein gravity
case corresponds to $-2\frac{\beta}{\gamma}$ by $-\frac{2\sin(\lambda \beta)}{\lambda\gamma}$. For example,
if the variables $(V,{\mathcal G},p_V,p_{\mathcal G})$ are used, we make the replacement  $p_V\rightarrow -\frac{2\sin(\lambda \beta)}{\lambda\gamma}$ in the Hamiltonian of Eq. (\ref{A5}).
%and
%if we consider the variables $(\bar{V},\bar{\phi},p_{\bar{V}},p_{\bar{\phi}})$, we will replace
%in  the Hamiltonian  (\ref{B8})
%$p_{\bar V}$ by $-\frac{2\sin(\lambda \beta)}{\lambda\gamma}$, because, in both cases, when
% $f(R)=R$ one has $p_V=p_{\bar{V}}=-2\frac{\beta}{\gamma}$.
It is conceivable that this way of introducing holonomy corrections, critically depends on the set of variables used to formulate the $F({\mathcal G})$ theory, which means
that the use of other canonically conjugate variables will lead to different corrections.
 %We can prove it, introducing holonomy corrections in Hamiltonians
 %(\ref{A5}) and (\ref{B8}), and showing that these corrections lead to different differential equations.
In the case that matter fluids are taken into account, considering the Hamiltonian (\ref{A5}) and its corresponding matter part, namely ${\mathcal H}_{matter}=\rho V$, upon
making the replacement,
$p_V\rightarrow -\frac{2\sin(\lambda \beta)}{\lambda\gamma}$, we finally obtain the following Hamiltonian,
\begin{eqnarray}\label{A12}
 {\mathcal H}_{hol}(V,{\mathcal G},p_V,p_{\mathcal G})
 =\frac{3}{V}\left(\frac{V^2p_{\mathcal G}}{4F''({\mathcal G})} \right)^{2/3}
 \nonumber\\+\frac{6\sin(\lambda \beta)}{\lambda\gamma}\left(\frac{V^2p_{\mathcal G}}{4F''({\mathcal G})} \right)^{1/3}+\frac{V}{2}\left({\mathcal G}F'({\mathcal G})-F({\mathcal G})\right)
 +\rho V.
\end{eqnarray}
The corresponding Hamilton equations are equal to,
\begin{eqnarray}\label{A13}
 \dot{V}=-\frac{\gamma}{2}\frac{\partial {\mathcal H}_{hol} }{\partial\beta};\quad \dot{\mathcal G}=\frac{\partial {\mathcal H}_{hol} }{\partial p_{\mathcal G}},
\end{eqnarray}
which together with the Hamiltonian constraint ${\mathcal H}_{hol}(V,{\mathcal G},p_V,p_{\mathcal G})=0$, lead to the following equations:
\begin{eqnarray}\label{A14}
 H=-\cos(\lambda\beta)\tilde{p}_{\mathcal G}^{\frac{1}{3}}\nonumber\\
 \dot{\mathcal G}=\frac{1}{2F''({\mathcal G})}\left(\tilde{p}_{\mathcal G}^{-\frac{1}{3}}+\tilde{p}_{\mathcal G}^{-\frac{2}{3}} \frac{\sin(\lambda \beta)}{\lambda\gamma} \right)
 \nonumber\\
 3\tilde{p}_{\mathcal G}^{\frac{2}{3}}+\tilde{p}_{\mathcal G}^{\frac{1}{3}} \frac{6\sin(\lambda \beta)}{\lambda\gamma}+\frac{1}{2}\left({\mathcal G}F'({\mathcal G})-F({\mathcal G})\right)
 +\rho=0,
\end{eqnarray}
where for notational simplicity, we introduced the variable $\tilde{p}_{\mathcal G}=\frac{p_{\mathcal G}}{4VF''({\mathcal G})}$.

Before dealing with these equations, we will check that in standard Einstein-Hilbert gravity where $F({\mathcal G})=0$, we can obtain the holonomy corrected Friedmann
equation of Eq. (\ref{A11}). In Einstein-Hilbert gravity, the
equations (\ref{A14}) take the following form,
\begin{eqnarray}\label{A15}
 H=-\cos(\lambda\beta)\tilde{p}_{\mathcal G}^{\frac{1}{3}}\nonumber\\
\tilde{p}_{\mathcal G}^{-\frac{1}{3}}+\tilde{p}_{\mathcal G}^{-\frac{2}{3}} \frac{\sin(\lambda \beta)}{\lambda\gamma}=0
 \nonumber\\
 3\tilde{p}_{\mathcal G}^{\frac{2}{3}}+\tilde{p}_{\mathcal G}^{\frac{1}{3}} \frac{6\sin(\lambda \beta)}{\lambda\gamma}
 +\rho=0.
\end{eqnarray}
The second and third equations lead to the relation $\tilde{p}_{\mathcal G}^{\frac{2}{3}}=\frac{\rho}{3}$, and the first and second equation can be written as follows,
\begin{eqnarray}\label{A16}
 \sin^2(\lambda\beta)=1-\frac{H^2}{\tilde{p}_{\mathcal G}^{\frac{2}{3}}}=1-\frac{3H^2}{\rho}\nonumber\\
\sin^2(\lambda\beta)=\frac{3}{\rho_c}\tilde{p}_{\mathcal G}^{\frac{2}{3}}=\frac{\rho}{\rho_c},
\end{eqnarray}
which eventually lead to the Friedmann equation (\ref{A11}) in LQC. Coming back to the general equation (\ref{A14}), these three equations have to be used to obtain a general relation of the form $G(H, {\mathcal G},\dot{\mathcal G},\rho)=0$ which will correspond to the modified  Friedmann equation in Gauss-Bonnet $F({\mathcal G})$ gravity. This equation in conjunction with the energy-momentum conservation equation $\dot{\rho}=-3H(\rho+P)$
and the equation of state $P=P(\rho)$, are the equations that will depict the dynamics of the system.

The combination of the first and third equations of (\ref{A14}) leads to the relation,
\begin{eqnarray}\label{A17}
12\rho_c(\tilde{p}_{\mathcal G}^{\frac{2}{3}}-H^2)=\left(\frac{1}{2}\left({\mathcal G}F'({\mathcal G})-F({\mathcal G})\right)+\rho +3 \tilde{p}_{\mathcal G}^{\frac{2}{3}} \right)^2,
\end{eqnarray}
which allows us to isolate $\tilde{p}_{\mathcal G}^{\frac{2}{3}} $ as a function of $ H$, $ {\mathcal G}$ and $\rho$, giving as a result
\begin{eqnarray}\label{A18}
\tilde{p}_{\mathcal G}^{\frac{2}{3}}=
\frac{4\rho_c-A({\mathcal G})}{6}\left(1-\sqrt{1-\frac{A^2({\mathcal G})+46\rho_c H^2}{(4\rho_c-A({\mathcal G}))^2}}  \right),
\end{eqnarray}
where the notation $A({\mathcal G})\equiv {\mathcal G}F'({\mathcal G})-F({\mathcal G})+2\rho $ has been introduced.

On the other hand, the combination of the three equations in (\ref{A14}) leads to the dynamical equation that corresponds to the modified Friedmann equation
in LQC-$F(\mathcal G)$ gravity,
\begin{eqnarray}\label{A19}
4\tilde{p}_{\mathcal G}^{\frac{4}{3}}(F''({\mathcal G}))^2 \dot{\mathcal G}^2
=\frac{\rho_c}{3}\left(1-\frac{H^2}{\tilde{p}_{\mathcal G}^{\frac{2}{3}}} \right)-\frac{1}{6}({\mathcal G}F'({\mathcal G})-F({\mathcal G}))-
\frac{\rho}{3}.
\end{eqnarray}
To sum up, the dynamical holonomy corrected equations in Gauss-Bonnet gravity are given by the modified Friedmann equation
in LQC-$F(\mathcal G)$ gravity and the conservation equation,
%the following equations,
\begin{eqnarray}\label{A20}
 4\tilde{p}_{\mathcal G}^{\frac{4}{3}}(F''({\mathcal G}))^2 \dot{\mathcal G}^2
=\frac{\rho_c}{3}\left(1-\frac{H^2}{\tilde{p}_{\mathcal G}^{\frac{2}{3}}} \right)-\frac{1}{6}({\mathcal G}F'({\mathcal G})-F({\mathcal G}))-
\frac{\rho}{3}\nonumber\\
\dot{\rho}=-3H(\rho+P),
%\nonumber \\
%\tilde{p}_{\mathcal G}^{\frac{2}{3}}=
%\frac{4\rho_c-A({\mathcal G})}{6}\left(1-\sqrt{1-\frac{A^2({\mathcal G})+46\rho_c H^2}{(4\rho_c-A({\mathcal G}))^2}}  \right),
\end{eqnarray}
where $\tilde{p}_{\mathcal G}^{\frac{2}{3}}$  also appears in Eq. (\ref{A18}) and $\rho_c$ is the critical density introduced at the end of the previous section.

Here, as in standard $F(\mathcal G)$ gravity, only two equations are needed to define the dynamics. All the Hamiltonian equations obtained from (\ref{A12}) are
equivalent to (\ref{A20}).

Note finally that, in order to solve Eq. (\ref{A20}), for a barotropic fluid an equation of state of the form
$P=P(\rho)$ is needed. In contrast, when one deals with a canonical scalar field $\varphi$, one has
$\rho=\frac{1}{2}\dot{\varphi}^2+V(\varphi)$ and $P=\frac{1}{2}\dot{\varphi}^2-V(\varphi)$.

\section{Analysis of the theory $F(\mathcal{G})$ with LQC}

In this way we shall investigate in which way we can have some LQC corrections in modified $F(\mathcal{G})$ gravity, at least the leading order LQC corrections. These corrections would materialize the first deviations from the classical non-LQC $F(\mathcal{G})$ gravity. In order to find these, we rewrite Eq. (\ref{A20}) in the following form,
\begin{equation}
24(F^{\prime\prime}(\mathcal{G})\Dot{\mathcal{G}})^2{\bar p}^2_{\mathcal{G}}(\mathcal{G})
-2\rho_c\left(1-\frac{H^2}{{\bar p}_{\mathcal{G}}(\mathcal{G})}\right)
+A(\mathcal{G})=0,
\label{fg1}
\end{equation}
where the parameter $\bar{p}_{\mathcal{G}}(\mathcal{G})$ stands for,
%$$A(\mathcal{G})=
%\mathcal{G}F^{\prime}(\mathcal{G})-F(\mathcal{G})+2\rho(t),$$
$$\bar{p}_{\mathcal{G}}(\mathcal{G})\equiv
\frac{1}{6}\left(4\rho_c-A(\mathcal{G})
-2\sqrt{2\rho_c}\sqrt{2\rho_c-A(\mathcal{G})-6H^2}\right).$$
Note that when $\rho_c\rightarrow\infty$, the quantity ${\bar p}_{\mathcal{G}}(\mathcal{G})$
can be represented as a series
$${\bar p}_{\mathcal{G}}(\mathcal{G})=
H^2+\frac{1}{48}\left(A(\mathcal{G})+6H^2\right)^2\frac{1}{\rho_c}
+\frac{1}{192}\left(A(\mathcal{G})+6H^2\right)^3\frac{1}{\rho_c^2}
+o\left(\frac{1}{\rho_c^3}\right).$$
Then, Eq. (\ref{fg1}) can be written as follows,
\begin{multline}
576H^6(F^{\prime\prime}(\mathcal{G})\Dot{\mathcal{G}})^2
-\left(A(\mathcal{G})-6H^2\right)^2
+\\+
\varepsilon\frac{\left(A(\mathcal{G})+6H^2\right)^2}{48H^2}
\left(A^2(\mathcal{G})-36H^4+1152H^6(F^{\prime\prime}(\mathcal{G})\Dot{\mathcal{G}})^2\right)+\\
+\varepsilon^2\frac{\left(A(\mathcal{G})+6H^2\right)^3}{2304H^4}\times \\  \times
\left(A^3(\mathcal{G})-6A^2(\mathcal{G})H^2+432H^6-576H^6(A(\mathcal{G}+30H^2))(F^{\prime\prime}(\mathcal{G})\Dot{\mathcal{G}})^2\right)+\ldots=0.
\label{fg2}
\end{multline}
It is easy to see that in the limit $\rho_c\rightarrow\infty$ this equation
reduced to the Friedman equation for $F(\mathcal{G})$ theory (see \cite{Fgg}). Therefore, we define the parameter $\varepsilon=1/\rho_c$, which as $\rho_c\rightarrow\infty$, it takes small values, and we seek a solution of this equation by performing a perturbative expansion in terms of this parameter.

In order to provide a consistent solution to the Cauchy problem or a boundary value problem, for the differential equation (\ref{fg1}), we shall perform a perturbative expansion in terms of the parameter $\varepsilon$, as $\varepsilon\rightarrow 0$. This expansion takes the form,
\begin{equation}
F(\mathcal{G})=\sum\limits_{k=0}^{\infty}\varepsilon^kF_k(\mathcal{G}).
\label{fg3}
\end{equation}
%Выражение (\ref{fg3}) подставляем в уравнение (\ref{fg2}).
%Затем члены уравнения (\ref{fg2}) разлагаем в ряд по малому параметру и собираем выражения
%при одинаковых степенях $\varepsilon$. Приравнивая полученные выражения
%(при одинаковых степенях $\varepsilon$) к нулю, приходим к системе уравнений для функций $F_k(G)$:
% Уравнение на F_0(G)
Then, the differential equations for each order of the expansion are given below,
\begin{equation}
24H^3F_0^{\prime\prime}(\mathcal{G})\Dot{\mathcal{G}}
-\mathcal{G}F_0^{\prime}(\mathcal{G})
+F_0(\mathcal{G})
+6H^2
-2\rho(t)=0,
\label{fg4}
\end{equation}
% Уравнение на F_1(G)
\begin{equation}
24H^3F_1^{\prime\prime}(\mathcal{G})\Dot{\mathcal{G}}
-\mathcal{G}F_1^{\prime}(\mathcal{G})
+F_1(\mathcal{G})=
-18H^4(1+2HF_0^{\prime\prime}(\mathcal{G})\Dot{\mathcal{G}})^2(1+6HF_0^{\prime\prime}
(\mathcal{G})\Dot{\mathcal{G}}),
\label{fg5}
\end{equation}
% Уравнение на F_2(G)
\begin{multline}
24H^3F_2^{\prime\prime}(\mathcal{G})\Dot{\mathcal{G}}
-\mathcal{G}F_2^{\prime}(\mathcal{G})
+F_2(\mathcal{G})=\\ =
-9H^5(3H(1+2HF_0^{\prime\prime}(\mathcal{G})\Dot{\mathcal{G}})^3
(4+39HF_0^{\prime\prime}(\mathcal{G})\Dot{\mathcal{G}}(1+2HF_0^{\prime\prime}(\mathcal{G})
\Dot{\mathcal{G}}))+\\
+4\Dot{\mathcal G}(1+2HF_0^{\prime\prime}(\mathcal{G})\Dot{\mathcal{G}})
(5+18HF_0^{\prime\prime}(\mathcal{G})\Dot{\mathcal{G}})F_1^{\prime\prime}(\mathcal{G})).
\label{fg6}
\end{multline}
where we included only the first three orders of the perturbative expansion of the solution. An important comment is in order: In principle, it is expected that the entire set of dynamical variables should be expanded in terms of the parameter $\varepsilon$, however if this is done for the Hubble rate, this would lead to a very restricted final form of the solutions. What is considered as a perturbative solution in the case at hand, is the full form of the $F(\mathcal{G})$, and the Hubble rate will be considered that it has a specific form. So we actually provide a perturbative method of reconstruction of the $F(\mathcal{G})$ gravity that generates such an evolution. In fact, the parameter $\rho_c$, and effectively $\varepsilon$, is justified only by the LQC context, and it is not expected to appear in the Hubble rate or in the matter energy density $\rho (t)$. The latter two physical quantities will be determined from the beginning, and in the later sections we shall investigate which $F(\mathcal{G})$ LQC-corrected gravity can generate the cosmological evolution corresponding to the given Hubble rate and matter energy density. Therefore, $\varepsilon$ enters the perturbative expansion only via Eq. (\ref{fg3}). If it is possible to construct a solution of equation (\ref{fg4}) for the function $F_0({\mathcal{G}})$,
the remaining terms of the expansion $F_k({\mathcal{G}})$, $k\geq 1$ are determined by solving linear differential equations. The initial or boundary conditions for the functions $F_k({\mathcal{G}})$ can be obtained by taking into account the expansion (\ref{fg3}) and the conditions for the equation (\ref{fg1}).

Note that from the equations defining the function $F_k({\mathcal{G}})$, it follows that the corrections to the $F_0({\mathcal{G}})$ are not related to the distribution of matter in the universe. In the following sections, we shall exemplify our results by using illustrative examples, emphasizing in bouncing cosmologies. We shall assume that all matter fluids are absent i.e.  $\rho(t)=0$, and therefore we are interested in vacuum modified LQC-corrected $F(\mathcal{G})$ theories.

\section{Power-law Cosmology from LQC-corrected $F(\mathcal{G})$ Gravity}

We start of our analysis, by studying a power-law cosmology in the context of LQC-corrected $F(\mathcal{G})$ gravity. Consider an evolution of the Universe for which the scale factor is of the following form,
$$a(t)=\alpha t^{2n},\quad \alpha\in\mathbb{R},\quad n\in\mathbb{N}.$$
For this model the Hubble rate reads,
$$H(t)=\frac{2n}{t},\quad {\mathcal{G}}(t)=\frac{192n^3(2n-1)}{t^4}.$$
Whence, it follows immediately that,
\begin{equation*}
H^2=\frac{{\mathcal{G}}^{1/2}n^{1/2}}{2\sqrt{3}\sqrt{2n-1}},\quad
H\Dot{{\mathcal{G}}}=-\frac{{\mathcal{G}}^{3/2}}{\sqrt{3}\sqrt{n}\sqrt{2n-1}}.
\end{equation*}
We assume that the initial conditions for the Cauchy problem of the differential equation (\ref{fg1}), are,
$$F({\mathcal{G}}_0)=\gamma_1,\quad F^{\prime}({\mathcal{G}}_0)=\gamma_2,$$
where $\gamma_1$, $\gamma_2$ are constants, $0<{\mathcal{G}}_0<+\infty$.
Then, for the system of differential equations for the functions $F_k({\mathcal{G}})$,
we obtain the following initial conditions:
$$F_0({\mathcal{G}}_0)=\gamma_1,\quad F_0^{\prime}({\mathcal{G}}_0)=\gamma_2,$$
\begin{equation}
F_k({\mathcal{G}}_0)=0,\quad F_k^{\prime}({\mathcal{G}}_0)=0,\quad k=1,2,\ldots.
\label{fg7}
\end{equation}

The general solution of the differential equation for the zero order unperturbed function $F_0(\mathcal{G})$, namely Eq. (\ref{fg4}), is of the following form:
$$F_0(\mathcal{G})=
c_1\mathcal{G}
+c_2\mathcal{G}^{(1-2n)/4}
-\frac{2\sqrt{3}\sqrt{\mathcal{G}}\sqrt{n(2n-1)}}{2n+1},$$
where $c_1$ and $c_2$ are arbitrary integration constants.

In order to further simplify our analysis, we can choose $\gamma_1$ and $\gamma_2$ in such a way,
so that $c_1=c_2=0$, without great loss of generality. In effect, $F_0(\mathcal{G})$ reads,
\begin{equation}
F_0({\mathcal{G}})=-\frac{2\sqrt{3}\sqrt{\mathcal{G}}\sqrt{n(2n-1)}}{2n+1}.
\label{fg9}
\end{equation}
Having $F_0({\mathcal{G}})$ at hand, we can easily proceed in finding the first order correction of the perturbative expansion,
namely the function $F_1({\mathcal{G}})$, which is of first order in the $\varepsilon$ expansion.
Considering Eqs. (\ref{fg7}) and (\ref{fg9}), the Cauchy problem for (\ref{fg5}) is the following:
$$\frac{4{\mathcal{G}}^2}{1-2n}F_1^{\prime\prime}({\mathcal{G}})-{\mathcal{G}}
F_1^{\prime}({\mathcal{G}})+F_1({\mathcal{G}})=
\frac{12{\mathcal{G}}n^3(n-1)}{(1-2n)(1+2n)^3},\quad
F_1({\mathcal{G}}_0)=0,\quad F_1^{\prime}({\mathcal{G}}_0)=0.$$
Therefore, one analytic solution of the Cauchy problem is:
$$F_1({\mathcal{G}})=c_3{\mathcal{G}}+c_4{\mathcal{G}}^{(1-2n)/4}
+\frac{12(1-n)n^3}{(1+2n)^3(3+2n)^2}\left(4{\mathcal{G}}-(3+2n){\mathcal{G}}\ln {\mathcal{G}}\right),$$
$$c_3=\frac{12(1-n)n^3\ln {\mathcal{G}}_0}{(1+2n)^3(3+2n)},\quad
c_4=-\frac{48(1-n)n^3{\mathcal{G}}_0^{(3+2n)/4}}{(1+2n)^3(3+2n)^2}.$$
Note that when $n=1$, the Cauchy problem has the trivial solution, $F_1({\mathcal{G}})=0$.
Iteratively, we can calculate the second order correction in the $\varepsilon$-expansion.
This can be easily done, since we have the explicit form of the function $F_1({\mathcal{G}})$ at hand.
Combining Eqs. (\ref{fg7}) and (\ref{fg9}) and also by taking into account the explicit form of the function
$F_1({\mathcal{G}})$, the Cauchy problem for the differential equation (\ref{fg6}), takes the form :
\begin{multline*}
\frac{4{\mathcal{G}}^2}{1-2n}F_2^{\prime\prime}({\mathcal{G}})-{\mathcal{G}}
F_2^{\prime}({\mathcal{G}})+F_2({\mathcal{G}})=\\
=c_3\frac{\sqrt{3}n^{3/2}(2n+3)(5n-2)}{4\sqrt{2n-1}(2n+1)^2}{\mathcal{G}}^{(3-2n)/4}
-\frac{3\sqrt{3}n^{9/2}(32n^3-78n^2+51n-20)}{(2n-1)^{3/2}(2n+1)^5(2n+3)}{\mathcal{G}}^{3/2},\\
F_2({\mathcal{G}}_0)=0,\quad F_2^{\prime}({\mathcal{G}}_0)=0.
\end{multline*}
It is easy to show that the solution to this problem has the following form,
\begin{multline*}
F_2({\mathcal{G}})=c_5{\mathcal{G}}+c_6{\mathcal{G}}^{(1-2n)/4}
+c_3\frac{\sqrt{3}n^{3/2}(20n^3+12n^2-23n+6)}{2\sqrt{2n-1}(2n+1)^3}{\mathcal{G}}^{(3-2n)/4}+\\
+\frac{6\sqrt{3}n^{9/2}(32n^3-78n^2+51n-20)}{\sqrt{2n-1}(2n+1)^5(2n+5)}{\mathcal{G}}^{3/2},
\end{multline*}
where $c_5$, $c_6$ are constants, which is easy to specify from the initial conditions given above. In principle, by continuing this iterative process, we can find all the higher order corrections of the perturbative $\varepsilon$-expansion.

In conclusion, for the case $n=1$, the resulting approximate solution of Eq. (\ref{fg1}),  has the following form,
\begin{equation}
F({\mathcal{G}})=-\frac{2\sqrt{{\mathcal{G}}}}{\sqrt{3}}
+\varepsilon^2\left(\frac{2\sqrt{{\mathcal{G}}_0}}{45\sqrt{3}}{\mathcal{G}}
-\frac{4{\mathcal{G}}_0^{7/4}}{315\sqrt{3}}\frac{1}{{\mathcal{G}}^{1/4}}
-\frac{2}{63\sqrt{3}}{\mathcal{G}}^{3/2}\right)+o(\varepsilon^2).
\label{fg9_sol}
\end{equation}
where we took into account only the first two corrections of the perturbative $\varepsilon$-expansion.
As a final task, we investigate the discrepancy between the approximate solution (\ref{fg9_sol}) and the one appearing in Eq. (\ref{fg1}).
For this purpose we substitute (\ref{fg9_sol}) in equation (\ref{fg1}) for this model
and as a result we obtain the function $g({\mathcal{G}},\varepsilon)=o(\varepsilon^2)$, $\varepsilon\rightarrow 0$. We have to note that the approximate solution is defined in the neighborhood of point ${\mathcal{G}}={\mathcal{G}}_0$.

\section{Bounce Cosmology from LQC $F({\mathcal G})$ Gravity}

One appealing cosmological scenario which is alternative to the standard inflationary description is the bounce
cosmology scenario \cite{bounce1,bounce2,bounce3,bounce4,bounce5,quintombounce,matterbounce,superbounce2,superbounce3}. According to this, the Universe
contracts until a minimal radius is reached, where it bounces off and starts to expand. In the bouncing cosmology context, the initial singularity is avoided
and this feature is mainly what renders the bouncing cosmologies so appealing, in comparison to the inflationary picture. Also, it is possible to consistently
describe early-time acceleration \cite{ekpyr1,ekpyr2,ekpyr3}, so in conjunction with the singularity avoidance, we have a very appealing candidate theory for our
Universe's evolution. In the context of modified gravity, bouncing cosmologies are easily realized \cite{superbounce2,superbounce3,superbounce4}, without the peculiar
requirements that the standard Einstein-Hilbert imposes in order for a bounce to occur \cite{bounce1}. In this
section we shall investigate which LQC-corrected $F({\mathcal G})$ gravities can realize bouncing cosmologies, using well known bouncing cosmologies paradigms. Special emphasis
shall be given on the exact form of the LQC-corrected $F({\mathcal G})$ which describes the bounce near the bouncing point. We address the problem of finding the
LQC-corrected $F({\mathcal G})$, using the perturbative expansion we used earlier.

\subsection{Exponential Bouncing Models}

We start off our analysis by studying some bouncing cosmologies with exponential scale factor. Before we get into the details of each model,
it is worth recalling in brief the conditions that define a bounce cosmology. For detailed accounts on these issues, see for example
\cite{bounce1,bounce2,bounce3,bounce4,bounce5,quintombounce}.

As we already mentioned, a cosmological bounce is described by two evolutionary eras, a
contraction and a subsequent expansion. In the contracting era, the Universe's scale factor $a(t)$ decreases up to a point, say at $t=t_s$, at which
a minimal radius is reached. This means that during the contraction, the derivative of the scale factor is negative $\dot{a}<0$, and at the minimal
radius point the derivative of the scale factor is zero, that is $\dot{a}=0$. At this point, the Universe bounces off and starts to expand, so that $\dot{a}>0$, until
a finite time singularity is probably reached. We shall study two bouncing models, the scale factor of which contains exponential functions of the cosmological time $t$.

\subsubsection{Symmetric Bounce Model}

\begin{figure}[h] \centering
\includegraphics[width=15pc]{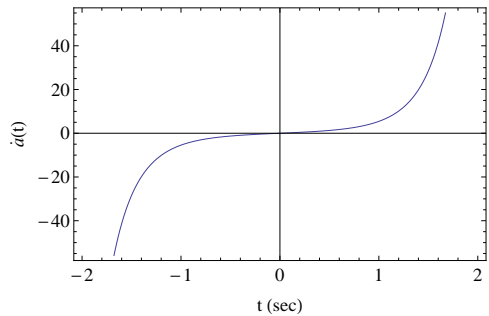}
\caption{The derivative of the scale factor $\dot{a}(t)$ as a function of the cosmic time $t$, for the bouncing cosmology $a(t)=e^{\alpha t^2}$, with $\alpha=1$.}
\label{plot1}
\end{figure}

In this section we study a symmetric bounce model, the scale factor of which is equal to,
\begin{equation}\label{scale}
a(t)=e^{\alpha t^2}\, ,
\end{equation}
where $\alpha$ is a positive arbitrary parameter. The scale factor (\ref{scale}) describes a cosmological bounce, for which the bouncing point is at $t=0$. Indeed, this can be verified in Fig. \ref{plot1}, where we plotted the time-dependence of the derivative of the scale factor $\dot{a}$. As it can be seen in Fig. \ref{plot1}, the function $\dot{a}$ is negative before the bouncing point $t=0$, equal to zero at the bouncing point and positive after the bouncing point.
What interests us the most in this section, is to investigate which LQC-corrected $F(G)$ theory can generate the bounce (\ref{scale}), near the bouncing point $t=0$. It is conceivable that what we need to examine is the limit $t\rightarrow 0$ of the differential equations (\ref{fg4}) and (\ref{fg5}) presented in the previous section. We shall be interested in finding the first two functions of the perturbative $\varepsilon$-expansion of the previous section, namely $F_0(\mathcal{G})$ and $F_1(\mathcal{G})$, assuming absence of any matter fluids. The Gauss-Bonnet invariant for the scale factor (\ref{scale}) reads,
\begin{equation}\label{gbscale}
\mathcal{G}=192 t^2 \alpha ^3 \left(1+2 t^2 \alpha \right)\, ,
\end{equation}
and in the limit $t\rightarrow 0$, this is approximately equal to,
\begin{equation}\label{gbscale1}
\mathcal{G}\simeq 192 t^2 \alpha ^3\, .
\end{equation}
Therefore, Eq. (\ref{gbscale1}) can be explicitly solved with respect to the cosmic time $t$,
\begin{equation}\label{coismictime}
t=\frac{\sqrt{\mathcal{G}}}{8 \sqrt{3} \alpha ^{3/2}}\,.
\end{equation}
Notice that the small $t$ limit is equivalent to the small $\mathcal{G}$ limit, so we shall take this observation into account in the rest of this section. In view of the above approximations, the differential equation that yields the function $F_0(\mathcal{G})$, namely Eq. (\ref{fg4}), at the vicinity of the bouncing point $t=0$ is equal to,
\begin{align}\label{diffeeqautaion1}
\frac{1}{24} \mathcal{G}^2 \left(48+\frac{\mathcal{G}}{\alpha ^2}\right)
%\frac{\mathrm{d}^2F_0(\mathcal{G})}{\mathrm{d}\mathcal{G}^2}-
F''_0(\mathcal{G})-
\mathcal{G}
%\frac{\mathrm{d}F_0(\mathcal{G})}{\mathrm{d}\mathcal{G}}
F'_0(\mathcal{G})
+F_0(\mathcal{G})+\frac{\mathcal{G}}{8 \alpha }=0\, ,
\end{align}
which can be solved analytically, with the solution being of the following form,
\begin{align}\label{so}
& F_0(\mathcal{G})=\mathcal{G} C_1-\frac{\sqrt{\mathcal{G}} \sqrt{48 \alpha ^2} C_2}{24 \alpha ^2} +\frac{-\mathcal{G} \ln[\mathcal{G}]+2 \sqrt{\mathcal{G}} \sqrt{48 \alpha ^2} \ln\left[\sqrt{48 \alpha ^2}\right]}{8 \alpha }\, .
\end{align}
where we kept only the leading order terms in the limit $\mathcal{G}\rightarrow 0$. Having $F_0(\mathcal{G})$ at hand, we can easily obtain the function $F_1(\mathcal{G})$, by solving the second differential equation given in Eq. (\ref{fg5}). In order to find an analytic approximation of the solution $F_1(\mathcal{G})$, we need to find an approximate form of the term,
\begin{equation}\label{ins}
\mathcal{C}=-18H^4(1+2HF_0^{\prime\prime}(\mathcal{G})\Dot{\mathcal{G}})^2(1+6HF_0^{\prime\prime}
(\mathcal{G})\Dot{\mathcal{G}})\, ,
\end{equation}
appearing in the $F_1(\mathcal{G})$ differential equation, where for notational simplicity we denoted the term with $\mathcal{C}$. The resulting expression of $\mathcal{C}$ is quite lengthy and can be found in the Appendix, so here we quote only the simplified expression for $\mathcal{C}$, where we will keep only the dominant terms in the limit $\mathcal{G}\rightarrow 0$. In this limit, the term $\mathcal{C}$ reads,
\begin{equation}\label{coeff}
\mathcal{C}=\mathcal{A}_1\sqrt{\mathcal{G}}\, ,
\end{equation}
where $\mathcal{A}_1$ stands for,
\begin{align}\label{dmathcal}
& \mathcal{A}_1=-\frac{C_2^3 \sqrt{\alpha ^2}}{128 \sqrt{3} \alpha ^3}+\frac{3 \sqrt{3} C_2^2 \sqrt{\alpha ^2} \ln\left[4 \sqrt{3} \sqrt{\alpha ^2}\right]}{64 \alpha ^2} -\frac{9 \sqrt{3} C_2 \sqrt{\alpha ^2} \ln\left[4 \sqrt{3} \sqrt{\alpha ^2}\right]^2}{32 \alpha }\\ \notag &+\frac{9}{16} \sqrt{3} \sqrt{\alpha ^2} \ln\left[4 \sqrt{3} \sqrt{\alpha ^2}\right]^3
\end{align}
Consequently, the differential equation that yields the solution for $F_1(\mathcal{G})$, namely Eq. (\ref{fg5}), becomes,
\begin{align}\label{diffeeqautaion1}
\frac{1}{24} \mathcal{G}^2 \left(48+\frac{\mathcal{G}}{\alpha ^2}\right)
%\frac{\mathrm{d}^2F_1(\mathcal{G})}{\mathrm{d}\mathcal{G}^2}-
F''_1(\mathcal{G})-
%\mathcal{G}\frac{\mathrm{d}F_1(\mathcal{G})}{\mathrm{d}\mathcal{G}}
\mathcal{G}F'_1(\mathcal{G})
+F_1(\mathcal{G})-\mathcal{A}_1\sqrt{\mathcal{G}}=0\, ,
\end{align}
which can analytically be solved to yield,
\begin{align}\label{f1sol}
& F_1(\mathcal{G})=e^{\frac{12 \left(-\alpha ^2+\sqrt{\alpha ^4}\right)}{\mathcal{G}}} \mathcal{G} C_1+\frac{e^{-\frac{12 \alpha ^2}{\mathcal{G}}-\frac{12 \sqrt{\alpha ^4}}{\mathcal{G}}} \mathcal{G} C_2}{24 \sqrt{\alpha ^4}}\, .
\end{align}
In conclusion, the final form of the small $\varepsilon$ limit of the $F(\mathcal{G})$, which generates the bounce (\ref{scale}) near the bouncing point, is approximately equal to,
\begin{align}\label{nearfinalform}
&F(\mathcal{G})\simeq F_0(\mathcal{G})+\varepsilon F_1(\mathcal{G})\simeq \\ & \notag
\mathcal{G} C_1-\frac{\sqrt{\mathcal{G}} \sqrt{48 \alpha ^2} C_2}{24 \alpha ^2} +\frac{-\mathcal{G} \ln[\mathcal{G}]+2 \sqrt{\mathcal{G}} \sqrt{48 \alpha ^2} \ln\left[\sqrt{48 \alpha ^2}\right]}{8 \alpha }\\ & \notag
\varepsilon \Big{(}e^{\frac{12 \left(-\alpha ^2+\sqrt{\alpha ^4}\right)}{\mathcal{G}}} \mathcal{G} C_1+\frac{e^{-\frac{12 \alpha ^2}{\mathcal{G}}-\frac{12 \sqrt{\alpha ^4}}{\mathcal{G}}} \mathcal{G} C_2}{24 \sqrt{\alpha ^4}}\Big{)}\,.
\end{align}
Recall that the variable $\varepsilon$ is equal to $1/\rho_c$, so practically the limit $\varepsilon\rightarrow 0$ measures the differences of the resulting LQC-corrected $F(\mathcal{G})$, in the limit where the LQC effects are strong, and therefore the LQC effects are contained in the term which is linear to $\varepsilon$. Indeed, the bounce is expected to occur when $\rho=\rho_c$, and at that point, $\mathcal{G}\rightarrow \infty$, and also, since $\rho_c$ is quite large, the parameter $\varepsilon$ is quite small, and in the expansion, the most dominant term of the perturbative expansion, is of the order $\sim \varepsilon$.

\subsubsection{Hyperbolic Cosine Bounce Model}

Another exponential bouncing model is described by the following scale factor,
$$a(t)=\cosh(\lambda t),\quad \lambda>0$$
For this model the Hubble rate is equal to,
$$H(t)=\lambda\tanh(\lambda t),\quad {\mathcal{G}}(t)=24\lambda^4\tanh^2(\lambda t).$$
Whence it follows immediately that,
\begin{equation*}
H^2=\frac{{\mathcal{G}}}{24\lambda^4},\quad H\Dot{{\mathcal{G}}}=2\lambda^2{\mathcal G}\left(1-\frac{{\mathcal{G}}}{24\lambda^4}\right).
\end{equation*}
Consider the following initial conditions for the Cauchy problem of Eq. (\ref{fg1}),
$$F({\mathcal{G}}_0)=\gamma_1,\quad F^{\prime}({\mathcal{G}}_0)=\gamma_2,$$
where $\gamma_1$, $\gamma_2$  are constants, and also $\mathcal{G}_0$ satisfies $0<{\mathcal{G}}_0<24\lambda^4$.
Then, for the functions $F_k({\mathcal{G}})$, the following boundary conditions hold true:
$$F_0({\mathcal{G}}_0)=\gamma_1,\quad F'_0({\mathcal{G}}_0)=\gamma_2,$$
\begin{equation}
F_k({\mathcal{G}}_0)=0,\quad F'_k({\mathcal{G}}_0)=0,\quad k=1,2,\ldots.
\label{fg10}
\end{equation}

The general solution of the unperturbed differential equation (\ref{fg4}) has the form:
\begin{multline}
F_0({\mathcal{G}})=c_1{\mathcal{G}}+c_2\sqrt{{\mathcal{G}}}\sqrt{24\lambda^4-{\mathcal{G}}}-\\
-\frac{1}{2\lambda^2}{\mathcal{G}}\ln\left(\frac{\sqrt{{\mathcal{G}}}}{2\sqrt{6}\lambda^2}\right)
+\frac{1}{\lambda^2}\sqrt{{\mathcal{G}}}\sqrt{24\lambda^4-{\mathcal{G}}}
\arctan\left(\frac{\sqrt{{\mathcal{G}}}+2\sqrt{6}\lambda^2}{\sqrt{24\lambda^4-{\mathcal{G}}}}\right),
\label{fg11}
\end{multline}
where $c_1$, $c_2$  are constants of integration. Note that this solution is defined for $0<\mathcal{G}<24\lambda^4$.
It is not difficult to find the values of the constants $c_1$ and $c_2$:
$$c_1=\gamma_1\frac{{\mathcal{G}}_0-12\lambda^4}{12\lambda^4{\mathcal{G}}_0}
+\gamma_2\frac{24\lambda^4-{\mathcal{G}}_0}{12\lambda^4}
+\frac{1}{2\lambda^2}\ln\left(\frac{\sqrt{{\mathcal{G}}_0}}{2\sqrt{6}\lambda^2}\right),$$
$$c_2=\gamma_1\frac{\sqrt{24\lambda^4-{\mathcal{G}}_0}}{12\lambda^4\sqrt{{\mathcal{G}}_0}}
-\gamma_2\frac{\sqrt{24\lambda^4-{\mathcal{G}}_0}\sqrt{{\mathcal{G}}_0}}{12\lambda^4}
-\frac{1}{\lambda^2}\arctan\left(\frac{2\sqrt{6}\lambda^2+
\sqrt{{\mathcal{G}}_0}}{\sqrt{24\lambda^4-{\mathcal{G}}_0}}\right).$$
Note that in the limit ${\mathcal{G}}_0\rightarrow 24\lambda^4$:
$$c_1\rightarrow \frac{\gamma_1}{24\lambda^4},\quad c_2\rightarrow-\frac{\pi}{2\lambda^2}.$$

One can consider the first-order correction of the perturbative $\varepsilon$-expansion, namely $F_1(\mathcal{G})$.
By considering Eqs. (\ref{fg10}) and (\ref{fg11}), the Cauchy problem for the differential equation (\ref{fg5}) becomes:
\begin{multline}
2{\mathcal{G}}^2\left(1-\frac{{\mathcal{G}}}{24\lambda^4}\right)F_1^{\prime\prime}({\mathcal{G}})-
{\mathcal{G}}F_1^{\prime}({\mathcal{G}})+F_1({\mathcal{G}})=\\
\frac{1}{2304\lambda^{10}}{\mathcal{G}}^2
({\mathcal{G}}({\mathcal{G}}-24\lambda^4)F_0^{\prime\prime}({\mathcal{G}})-6\lambda^2)^2
({\mathcal{G}}({\mathcal{G}}-24\lambda^4)F_0^{\prime\prime}({\mathcal{G}})-2\lambda^2),
\label{fg12}
\end{multline}
\begin{equation}
F_1({\mathcal{G}}_0)=0,\quad F'_1({\mathcal{G}}_0)=0.
\label{fg13}
\end{equation}
It is easy to show that the solutions of the homogeneous differential equation are:
$$y_1({\mathcal{G}})={\mathcal{G}},\quad y_2({\mathcal{G}})=\sqrt{{\mathcal{G}}}\sqrt{24\lambda^4-{\mathcal{G}}}.$$
Then the solution of the inhomogeneous equation (\ref{fg12}) can be written as:
\begin{multline*}
F_1({\mathcal{G}})=c_3{\mathcal G}+c_4\sqrt{{\mathcal{G}}}\sqrt{24\lambda^4-{\mathcal{G}}}+\\
+\frac{{\mathcal{G}}}{2304\lambda^{10}}
\int\limits({\mathcal{G}}({\mathcal{G}}-24\lambda^4)F_0^{\prime\prime}({\mathcal{G}})-2\lambda^2)
({\mathcal{G}}({\mathcal{G}}-24\lambda^4)F_0^{\prime\prime}({\mathcal{G}})-6\lambda^2)^2d{\mathcal{G}}-\\
-\frac{\sqrt{{\mathcal{G}}}\sqrt{24\lambda^4-{\mathcal{G}}}}{2304\lambda^{10}}
\int\limits\frac{\sqrt{{\mathcal{G}}}}{\sqrt{24\lambda^4-{\mathcal{G}}}}
({\mathcal{G}}({\mathcal{G}}-24\lambda^4)F_0^{\prime\prime}({\mathcal{G}})-2\lambda^2)
({\mathcal{G}}({\mathcal{G}}-24\lambda^4)F_0^{\prime\prime}({\mathcal{G}})-6\lambda^2)^2d{\mathcal{G}},
\end{multline*}
where $c_3$, $c_4$ are constants of integration. We can explicitly calculate the integral by using a new variable,
$$s=\arctan\left(\frac{\sqrt{{\mathcal{G}}}+2\sqrt{6}\lambda^2}{\sqrt{24\lambda^4-{\mathcal{G}}}}\right),
\quad \frac{\pi}{4}\leq s<\frac{\pi}{2}.$$
Then, the first correction $F_1({\mathcal{G}})$ can be written as follows,
\begin{multline}
F_1(s)=12\lambda^4(2c_3\cos^22s-c_4\sin 4s)+\\
+\frac{9}{4}\lambda^4\left(1+4s^2+76(s+c_2\lambda^2)^2+72c_2^2\lambda^4\ln(24\lambda^4)\right)\cos^22s-\\
-\frac{9}{4}c_2\lambda^6\left(9\ln 2+11\ln(24\lambda^4)\right)\sin 4s
-\frac{9}{8}\lambda^4\arctan(\cot 2s)(8c_2\lambda^2\cos^22s+\sin 4s)+\\
+18\lambda^4\left(4(7\mathcal{A}^{\prime}(s)+2\mathcal{B}^{\prime}(s))\cos^22s
+(18\mathcal{A}(s)-18\mathcal{B}(s)+5\mathcal{A}^{\prime\prime}(s))\sin 4s\right)-\\
-18\lambda^4(s+c_2\lambda^2)\left((18\mathcal{A}^{\prime}(s)-18\mathcal{B}^{\prime}(s)+5\ln(\sin 2s))\sin 4s
+4(7\mathcal{A}^{\prime\prime}(s)+2\mathcal{B}^{\prime\prime}(s))\cos^22s\right)+\\
+18\lambda^4(s+c_2\lambda^2)^2\left(9(\mathcal{A}^{\prime\prime}(s)-\mathcal{B}^{\prime\prime}(s))\sin 4s
+2(2\ln|\cos 2s|+7\ln\sin 2s)\cos^22s\right)-\\
-54\lambda^4(s+c_2\lambda^2)^3\left(\cos 6s\csc 2s+\sin 4s\ln|\tan 2s|\right),
\label{fg14}
\end{multline}
where the functions $\mathcal{A}(s)$ and $\mathcal{B}(s)$ are defined as follows:
$$\mathcal{A}(s)=-\frac{11}{36}s^3+\frac{1}{6}s^3\ln 2s
+\sum\limits_{n=1}^{\infty}(-1)^n\frac{2^{4n-1}}{k(2k+3)!}B_{2k}s^{2k+3},$$
$$\mathcal{B}(s)=\sum\limits_{n=1}^{\infty}(-1)^n\frac{2^{2n-4}(1-2^{2k})}{k(2k+3)!}B_{2k}(\pi-2s)^{2k+3},$$
and $B_{2k}$ are the Bernoulli numbers.
Note that
$$\mathcal{A}^{\prime\prime\prime}(s)=\ln\sin 2s,\quad \mathcal{B}^{\prime\prime\prime}(s)=\ln|\cos 2s|.$$

From the initial conditions (\ref{fg13}), it is easy to find the coefficients $c_3$ and $c_4$, but
due to lengthy and complicated expressions, we do not present the final solution to the Cauchy problem.

Before closing this section, we need to note that the following limiting cases hold true,
$$\lim\limits_{{\mathcal{G}}\rightarrow 0^{+}}F_0({\mathcal{G}})=0,\quad
\lim\limits_{{\mathcal{G}}\rightarrow 24\lambda^4{}^{-}}F_0({\mathcal{G}})=24c_1\lambda^4.$$
$$\lim\limits_{{\mathcal{G}}\rightarrow 0^{+}}F_1({\mathcal{G}})=0,\quad
\lim\limits_{{\mathcal{G}}\rightarrow 24\lambda^4{}^{-}}F_1({\mathcal{G}})=
\begin{cases}
\frac{9}{4}\lambda^4\left(1+18\pi^2\ln(24\lambda^4)
+224\mathcal{A}^{\prime}\left(\frac{\pi}{2}\right)\right),& c_2=-\frac{\pi}{2\lambda^2};\\
+\infty,& c_2\neq-\frac{\pi}{2\lambda^2}.
\end{cases}$$
This implies that for the case $c_2\neq-\frac{\pi}{2\lambda^2}$, the first order correction $F_1({\mathcal{G}})$ is defined only in a neighborhood of the point ${\mathcal{G}}={\mathcal{G}}_0$. However, the case $c_2=-\frac{\pi}{2\lambda^2}$ should be considered a separate way. It can be obtained, for example, in the formulation of the boundary value problem of Eq. (\ref{fg1}):
$$\lim\limits_{{\mathcal{G}}\rightarrow 0^{+}}F({\mathcal{G}})=0,\quad \lim\limits_{{\mathcal{G}}\rightarrow 24\lambda^4{}^{-}}F({\mathcal{G}})=0.$$
At the same time it is easy to show that,
$$c_1=0,\quad c_2=-\frac{\pi}{2\lambda^2},\quad
c_3=-\frac{3}{32}\left(1+18\pi^2\ln(24\lambda^4)+224\mathcal{A}^{\prime}\left(\frac{\pi}{2}\right)\right),\quad
c_4\in\mathbb{R}.$$
Note also that in this case, the first order correction $F_1({\mathcal{G}})$ can take finite values on the whole interval,
where the function $F_0({\mathcal{G}})$  is defined, which is the following interval,
$$\left|\frac{F_1({\mathcal{G}})}{F_0({\mathcal{G}})}\right|<C,\quad 0\leq {\mathcal{G}}\leq 24\lambda^4,$$
where $C$ is a finite constant.

Similarly, one can build  corrections any other order for the model $a(t)=\cosh(\lambda t)$, but the expressions can be more complicated so we confine ourselves to the first two orders.

\subsection{Power-law Bouncing Model}

\begin{figure}[h] \centering
\includegraphics[width=15pc]{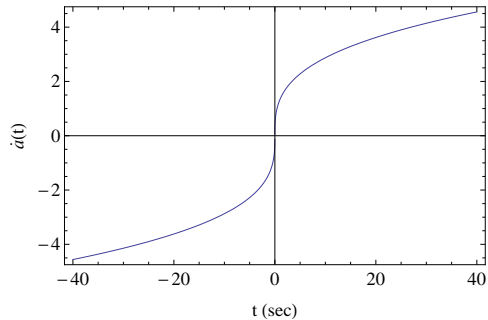}
\caption{The derivative of the scale factor $\dot{a}(t)$ as a function of the cosmic time $t$, for the bouncing cosmology $a(t)=(t-t_s)^{\alpha}$, with $\alpha=4/3$ and $t_s=10^{-35}$sec.}
\label{plot2}
\end{figure}

As a final example, we shall study another bouncing cosmology which is described by the following scale factor,
\begin{equation}\label{scale1bounce}
a(t)=(t-t_s)^{\alpha}\, ,
\end{equation}
which is related to certain ekpyrotic models \cite{superbounce2,superbounce3}. The parameter $\alpha$ is a real positive number which for the purposes of this section we choose it to satisfy $1<\alpha<5$, for reasons to be clear later on in this section. The cosmological evolution described by the scale factor (\ref{scale1bounce}) perfectly describes a bounce, meaning that before the bouncing point, we have $\dot{a}<0$, after the bouncing point $\dot{a}>0$ and at the bounce $\dot{a}=0$, as can easily be checked by looking at Fig. \ref{plot2}, where we plotted the time dependence of the function $\dot{a}$, for $\alpha=4/3$ \footnote{Notice that in principle if $\alpha$ is not appropriately chosen, the scale factor might turn complex, so extra attention should be payed on this issue.} and $t_s=10^{-35}$sec.
Since for general values of $\alpha$, it is quite difficult to analytically solve the differential equations (\ref{fg5}), as in the symmetric bounce case, we shall
investigate here which LQC-corrected $F({\mathcal G})$ gravity can describe the bounce (\ref{scale1bounce}) near the bouncing point $t=t_s$, which can be arbitrarily chosen. Notice that
when $t\simeq t_s$, the expression $x=t-t_s$ tends to zero, and this observation shall be useful in the following analysis. The Gauss-Bonnet invariant ${\mathcal G}$, for
the scale factor (\ref{scale1bounce}) reads,
\begin{equation}\label{gnbinv}
\mathcal{G}=\frac{24 (-1+\alpha ) \alpha ^3}{(t-t_s)^4}\, ,
\end{equation}
so as $t\rightarrow t_s$, the variable $\mathcal{G}$ increases. Therefore the limit $x\rightarrow 0$ corresponds to the limit $\mathcal{G}\rightarrow \infty$, and we shall
use this correspondence in the analysis that follows. Solving Eq. (\ref{gnbinv}), with respect to $t-t_s$, we obtain,
\begin{equation}\label{shdge}
t-t_s=\frac{2^{3/4} 3^{1/4} \left(-\alpha ^3+\alpha ^4\right)^{1/4}}{\mathcal{G}^{1/4}}\, ,
\end{equation}
and since $1<\alpha<5$, no inconsistency related to complex cosmological times occurs. The differential equation that yields the $F_0(\mathcal{G})$ gravity is therefore equal to,
\begin{align}\label{diffeeqautaion234}
\mathcal{B}_1 \mathcal{G}^2
%\frac{\mathrm{d}^2F_0(\mathcal{G})}{\mathrm{d}\mathcal{G}^2}-
F''_0(\mathcal{G})-
\mathcal{G}
F'_0(\mathcal{G})
%\frac{\mathrm{d}F_0(\mathcal{G})}{\mathrm{d}\mathcal{G}}
+F_0(\mathcal{G})+\mathcal{D}\sqrt{\mathcal{G}}=0\, ,
\end{align}
where we have set for simplicity $\mathcal{D}$ and $\mathcal{B}_1$ to be equal to,
\begin{equation}\label{arce}
\mathcal{D}=\frac{\sqrt{\frac{3}{2}} \alpha ^2}{\sqrt{(-1+\alpha ) \alpha ^3}}\,{\,}{\,}{\,}\mathcal{B}_1=-\frac{4}{\alpha-1} .
\end{equation}
The solution to the differential equation (\ref{diffeeqautaion234}) is equal to,
\begin{align}\label{so111}
& F_0(\mathcal{G})=\frac{4 \mathcal{D} \sqrt{\mathcal{G}}}{-2+\mathcal{B}_1}+\mathcal{G}^{\frac{1}{\mathcal{B}_1}} C_1+\mathcal{G} C_2\, .
\end{align}
Finally, we calculate the first order correction of the LQC-corrected $F(\mathcal{G})$ gravity, namely the function $F_1(\mathcal{G})$, which easily follows if we use the analytic form of $F_0(\mathcal{G})$ given in Eq. (\ref{so111}). However, since the exact solution is quite complicated and lengthy, we can approximate the resulting expression by recalling that the limit $x\rightarrow 0$, corresponds to $\mathcal{G}\rightarrow \infty$, so the larger term dominates in the expressions. For $1<\alpha<5$, the most dominant term is of the order $\sim \mathcal{G}$, so $F(\mathcal{G})$ is approximately equal to $F_0(\mathcal{G})\simeq C_2 \mathcal{G} $, and the resulting differential equation that yields the solution for $F_1(\mathcal{G})$, namely Eq. (\ref{fg5}), becomes,
\begin{align}\label{diffeeqautaion111111111}
\mathcal{B}_1 \mathcal{G}^2
%\frac{\mathrm{d}^2F_1(\mathcal{G})}{\mathrm{d}\mathcal{G}^2}
F''_1(\mathcal{G})
-\mathcal{G}
%\frac{\mathrm{d}F_1(\mathcal{G})}{\mathrm{d}\mathcal{G}}
F'_1(\mathcal{G})
+F_1(\mathcal{G})-\frac{3 \mathcal{G} \alpha ^4}{4 \left(-\alpha ^3+\alpha ^4\right)}=0\, ,
\end{align}
which can analytically be solved to yield,
\begin{align}\label{f1solsq}
& F_1(\mathcal{G})=\mathcal{G}^{\frac{1}{\mathcal{B}_1}} C_1+\mathcal{G} C_2+\frac{3 \mathcal{G} \alpha  (-\mathcal{B}_1-\ln[\mathcal{G}]+\mathcal{B}_1 \ln[\mathcal{G}])}{4 (-1+\mathcal{B}_1)^2 (-1+\alpha )}\, .
\end{align}

\section{Conclusions}

In this paper, we extended the holonomy corrections formalism of LQC to Gauss-Bonnet $F(\mathcal{G})$ modified gravity theories. Specifically, upon using the method of Lagrange multipliers, we constructed the classical dynamical cosmological equations in Gauss-Bonnet gravity, for a flat FLRW geometry. In addition, we extended LQC to Gauss-Bonnet gravity obtaining a holonomy corrected Friedmann equation, which contains all the dynamical information of the system. Then after explaining the reconstruction method in holonomy corrected Gauss-Bonnet gravity, we applied our formalism to certain cosmological scenarios, focusing in the realization of these cosmologies in the context of LQC-corrected $F(\mathcal{G})$ gravity. The cosmological scenarios on which we emphasized are the bouncing cosmologies and the reason for that is that bouncing cosmologies provide a quite elegant alternative scenario to the inflationary paradigm. As we evinced, the resulting LQC-corrected $F(\mathcal{G})$ dynamical equations are quite
complicated, so we performed a perturbative expansion, using as a perturbation parameter, the parameter $\varepsilon$ which is related to a very important physical quantity, the critical density $\rho_c$, as $\varepsilon=1/\rho_c$. Practically, the parameter $\rho_c$ measures how quantum is the theory, and the smaller it is, the theory ``stretches'' in the more quantum era. In the perturbation series we used, we assumed that $\varepsilon\rightarrow 0$, so the leading order corrections we found, practically quantify the way that the LQC-corrected theory deviates from the classical $F(\mathcal{G})$ theory. So iteratively, one can recover higher and higher corrections, being though less significant, depending on how fast $\rho_c$ tends to infinity.

Moreover, the possibility of having finite time singularities
\cite{sergeinojiri} and specifically mild types of singularities
\cite{sno2015} is also quite interesting and should be extensively
discussed.

Finally, the viability of the theories should also be checked and confronted with current observational data,
but this task extends the purpose of this paper, which was the presentation of the method of LQC-corrected $F(\mathcal{G})$ gravity,
and its usefulness towards realizing cosmological scenarios. We hope however to discuss these topics in the near future.

\vspace{1cm}
{\bf Acknowledgments:}
This investigation has been
supported in part by MINECO (Spain)
projects MTM2011-27739-C04-01 (J.H.),  FIS2010-15640 and FIS2013-44881 (S.D.O.) and Russian Ministry of Education and Science (S.D.O. and A.N.M.).

%{\bf HERE YOU HAVE TO
%INTRODUCE ALL THE PROJECTS WHERE YOU PARTICIPATE.}

\section*{Appendix}

In this appendix, we give the exact form of the parameter $\mathcal{C}$ appearing in Eq. (\ref{ins}). It's detailed form is,
\begin{align}\label{cparamdetailed}
& \mathcal{C}=\frac{C_2^2 \mathcal{G}^4}{4718592 \alpha ^8}+\frac{\mathcal{G}^5}{4718592 \alpha ^8}+\frac{5 C_2^2 \mathcal{G}^3}{221184 \alpha ^6} +\frac{\mathcal{G}^4}{147456 \alpha ^6}+\frac{13 C_2^2 \mathcal{G}^2}{18432 \alpha ^4}+\frac{C_2^2 \mathcal{G}}{192 \alpha ^2}\\ & \notag-\frac{C_2^3 \mathcal{G}^{7/2} \sqrt{\alpha ^2}}{14155776 \sqrt{3} \alpha ^9}  -\frac{C_2 \mathcal{G}^{9/2} \sqrt{\alpha ^2}}{1572864 \sqrt{3} \alpha ^9}-\frac{C_2^3 \mathcal{G}^{5/2} \sqrt{\alpha ^2}}{98304 \sqrt{3} \alpha ^7} -\frac{13 C_2 \mathcal{G}^{7/2} \sqrt{\alpha ^2}}{294912 \sqrt{3} \alpha ^7}-\frac{C_2^3 \mathcal{G}^{3/2} \sqrt{\alpha ^2}}{2048 \sqrt{3} \alpha ^5}\\ & \notag-\frac{C_2 \mathcal{G}^{5/2} \sqrt{\alpha ^2}}{1536 \sqrt{3} \alpha ^5} -\frac{C_2^3 \sqrt{\mathcal{G}} \sqrt{\alpha ^2}}{128 \sqrt{3} \alpha ^3}-\frac{C_2 \mathcal{G}^4 \ln\left[4 \sqrt{3} \sqrt{\alpha ^2}\right]}{393216 \alpha ^7} -\frac{5 C_2 \mathcal{G}^3 \ln\left[4 \sqrt{3} \sqrt{\alpha ^2}\right]}{18432 \alpha ^5}
\\ & \notag-\frac{13 C_2 \mathcal{G}^2 \ln\left[4 \sqrt{3} \sqrt{\alpha ^2}\right]}{1536 \alpha ^3}-\frac{C_2 \mathcal{G} \ln\left[4 \sqrt{3} \sqrt{\alpha ^2}\right]}{16 \alpha }+\frac{C_2^2 \mathcal{G}^{7/2} \sqrt{\alpha ^2} \ln\left[4 \sqrt{3} \sqrt{\alpha ^2}\right]}{786432 \sqrt{3} \alpha ^8}\\ & \notag +\frac{\mathcal{G}^{9/2} \sqrt{\alpha ^2} \ln\left[4 \sqrt{3} \sqrt{\alpha ^2}\right]}{262144 \sqrt{3} \alpha ^8}+\frac{\sqrt{3} C_2^2 \mathcal{G}^{5/2} \sqrt{\alpha ^2} \ln\left[4 \sqrt{3} \sqrt{\alpha ^2}\right]}{16384 \alpha ^6}\\ & \notag +\frac{13 \mathcal{G}^{7/2} \sqrt{\alpha ^2} \ln\left[4 \sqrt{3} \sqrt{\alpha ^2}\right]}{49152 \sqrt{3} \alpha ^6}+\frac{3 \sqrt{3} C_2^2 \mathcal{G}^{3/2} \sqrt{\alpha ^2} \ln\left[4 \sqrt{3} \sqrt{\alpha ^2}\right]}{1024 \alpha ^4}\\ & \notag +\frac{\mathcal{G}^{5/2} \sqrt{\alpha ^2} \ln\left[4 \sqrt{3} \sqrt{\alpha ^2}\right]}{256 \sqrt{3} \alpha ^4}+\frac{3 \sqrt{3} C_2^2 \sqrt{\mathcal{G}} \sqrt{\alpha ^2} \ln\left[4 \sqrt{3} \sqrt{\alpha ^2}\right]}{64 \alpha ^2}+\frac{3}{16} \mathcal{G} \ln\left[4 \sqrt{3} \sqrt{\alpha ^2}\right]^2\\ & \notag +\frac{\mathcal{G}^4 \ln\left[4 \sqrt{3} \sqrt{\alpha ^2}\right]^2}{131072 \alpha ^6}+\frac{5 \mathcal{G}^3 \ln\left[4 \sqrt{3} \sqrt{\alpha ^2}\right]^2}{6144 \alpha ^4}+\frac{13 \mathcal{G}^2 \ln\left[4 \sqrt{3} \sqrt{\alpha ^2}\right]^2}{512 \alpha ^2}\\ & \notag -\frac{C_2 \mathcal{G}^{7/2} \sqrt{\alpha ^2} \ln\left[4 \sqrt{3} \sqrt{\alpha ^2}\right]^2}{131072 \sqrt{3} \alpha ^7}-\frac{3 \sqrt{3} C_2 \mathcal{G}^{5/2} \sqrt{\alpha ^2} \ln\left[4 \sqrt{3} \sqrt{\alpha ^2}\right]^2}{8192 \alpha ^5}-\frac{9 \sqrt{3} C_2 \mathcal{G}^{3/2} \sqrt{\alpha ^2} \ln\left[4 \sqrt{3} \sqrt{\alpha ^2}\right]^2}{512 \alpha ^3}\\ & \notag -\frac{9 \sqrt{3} C_2 \sqrt{\mathcal{G}} \sqrt{\alpha ^2} \ln\left[4 \sqrt{3} \sqrt{\alpha ^2}\right]^2}{32 \alpha }+\frac{9}{16} \sqrt{3} \sqrt{\mathcal{G}} \sqrt{\alpha ^2} \ln\left[4 \sqrt{3} \sqrt{\alpha ^2}\right]^3+\frac{\mathcal{G}^{7/2} \sqrt{\alpha ^2} \ln\left[4 \sqrt{3} \sqrt{\alpha ^2}\right]^3}{65536 \sqrt{3} \alpha ^6}\\ & \notag +\frac{3 \sqrt{3} \mathcal{G}^{5/2} \sqrt{\alpha ^2} \ln\left[4 \sqrt{3} \sqrt{\alpha ^2}\right]^3}{4096 \alpha ^4}+\frac{9 \sqrt{3} \mathcal{G}^{3/2} \sqrt{\alpha ^2} \ln\left[4 \sqrt{3} \sqrt{\alpha ^2}\right]^3}{256 \alpha ^2}
\end{align}
and consequently, in the small $\mathcal{G}$ limit, it can be approximated by Eq. (\ref{coeff}).


\begin{thebibliography}{0}



\bibitem{penrose}

R.~Penrose,
  %``Gravitational collapse: The role of general relativity,''
  [Gen.\ Rel.\ Grav.\  {\bf 34}, 1141 (2002)].
  %%CITATION = RNCIB,1,252;%%
  %644 citations counted in INSPIRE as of 22 Apr 2015

\bibitem{LQC}
A.~Ashtekar and A.~Barrau,
  %``Loop quantum cosmology: From pre-inflationary dynamics to observations,''
  arXiv:1504.07559 [gr-qc];\\
B.~Bolliet, J.~Grain, C.~Stahl, L.~Linsefors and A.~Barrau,
  %``Comparison of primordial tensor power spectra from the deformed algebra and dressed metric approaches in loop quantum cosmology,''
  Phys.\ Rev.\ D {\bf 91} (2015) 8,  084035
  [arXiv:1502.02431 [gr-qc]];\\
L.~C.~Gomar, M.~Martín-Benito and G.~A.~M.~Marugán,
  %``Gauge-Invariant Perturbations in Hybrid Quantum Cosmology,''
  JCAP {\bf 1506} (2015) 06,  045
  [arXiv:1503.03907 [gr-qc]];\\
 A.~Barrau, M.~Bojowald, G.~Calcagni, J.~Grain and M.~Kagan,
  %``Anomaly-free cosmological perturbations in effective canonical quantum gravity,''
  JCAP {\bf 1505} (2015) 05,  051
  [arXiv:1404.1018 [gr-qc]].;\\
I.~Agullo, A.~Ashtekar and W.~Nelson,
  %``A Quantum Gravity Extension of the Inflationary Scenario,''
  Phys.\ Rev.\ Lett.\  {\bf 109} (2012) 251301
  [arXiv:1209.1609 [gr-qc]].;\\
%\cite{Ashtekar:2011ni}
%\bibitem{Ashtekar:2011ni}
A.~Ashtekar and P.~Singh,
%``Loop Quantum Cosmology: A Status Report,''
Class.\ Quant.\ Grav.\  {\bf 28} (2011) 213001
[arXiv:1108.0893 [gr-qc]]; \\
%%CITATION = ARXIV:1108.0893;%%
%222 citations counted in INSPIRE as of 14 Feb 2015
%\cite{Ashtekar:2007tv}
%\bibitem{Ashtekar:2007tv}
A.~Ashtekar,
%``An Introduction to Loop Quantum Gravity Through Cosmology,''
Nuovo Cim.\ B {\bf 122} (2007) 135
[gr-qc/0702030]; \\
%%CITATION = GR-QC/0702030;%%
%112 citations counted in INSPIRE as of 14 Feb 2015
%\cite{Bojowald:2008ik}
%\bibitem{Bojowald:2008ik}
M.~Bojowald,
%``Consistent Loop Quantum Cosmology,''
Class.\ Quant.\ Grav.\  {\bf 26} (2009) 075020
[arXiv:0811.4129 [gr-qc]]; \\
%%CITATION = ARXIV:0811.4129;%%
%47 citations counted in INSPIRE as of 14 Feb 2015
%\cite{Cailleteau:2012fy}
%\bibitem{Cailleteau:2012fy}
T.~Cailleteau, A.~Barrau, J.~Grain and F.~Vidotto,
%``Consistency of holonomy-corrected scalar, vector and tensor perturbations in Loop Quantum Cosmology,''
Phys.\ Rev.\ D {\bf 86} (2012) 087301
[arXiv:1206.6736 [gr-qc]];\\
%%CITATION = ARXIV:1206.6736;%%
%35 citations counted in INSPIRE as of 14 Feb 2015
%\cite{Quintin:2014oea}
%\bibitem{Quintin:2014oea}
A. Ashtekar, P. Singh, Class. Quant. Grav. 28, 213001 (2011) [arXiv:1108.0893 ];\\
A. Corichi, P. Singh, Phys.Rev. D80 (2009) 044024 [arXiv:0905.4949];\\
P. Singh, Class.Quant.Grav. 26 (2009) 125005 [arXiv:0901.2750];\\
A. Ashtekar, T. Pawlowski, P. Singh, Phys.Rev. D74 (2006) 084003 [gr-qc/0607039];\\
M. Bojowald, Class.Quant.Grav. 26 (2009) 075020 [arXiv:0811.4129]




\bibitem{zm11}
X. Zhang and Y. Ma,  Phys. Rev. Lett. {\bf 106}, 171301 (2011)
  [arXiv:1101.1752].
\bibitem{zm11a}
X. Zhang and Y. Ma,  Phys. Rev. {\bf D84}, 064040 (2011)
  [arXiv:1107.4921].\\
%\bibitem{zm13}
X. Zhang and Y. Ma,  Front. Phys.  {\bf 8}, 80 (2013)
  [arXiv:1211.5024].


\bibitem{Risi}
G. De Risi, R. Maartens and P. Singh,  Phys. Rev.  {\bf D76}, 103531 (2007)
  [arXiv:0706.3586].

\bibitem{aho} J. Amor\'os, J. de Haro and S.D. Odintsov Phys. Rev.  {\bf D89}, 104010 (2014)
[arXiv:1402.3071].

\bibitem{haro14} J. de Haro, EPL {\bf 107}, 29001 (2014)
 [arXiv:1403.4529].

\bibitem{bmmo} K. Bamba, A.N. Makarenko, A.N. Myagki and S.D. Odintsov, Phys. Lett. {\bf B732}, 349 (2014)
 [arXiv:1403.3242].


\bibitem{bounces}
M.~Novello and S.~E.~P.~Bergliaffa,
  Phys.\ Rept.\  {\bf 463}, 127 (2008)
  [arXiv:0802.1634].\\
  J.~L.~Lehners,
  Phys.\ Rept.\  {\bf 465}, 223 (2008)
  [arXiv:0806.1245].\\
  R.~H.~Brandenberger,
  Int.\ J.\ Mod.\ Phys.\ Conf.\ Ser.\  {\bf 01}, 67 (2011)
  [arXiv:0902.4731]. \\
  R.~H.~Brandenberger,
  AIP Conf.\ Proc.\  {\bf 1268}, 3 (2010)
  [arXiv:1003.1745].\\
  D.~Battefeld and P.~Peter,
  [arXiv:1406.2790].

\bibitem{bounce1} M. Novello, S.E.Perez Bergliaffa, Phys.Rept. {\bf 463} (2008) 127 [arXiv:0802.1634];\\
%\cite{Li:2014era}
%\bibitem{Li:2014era}
  C.~Li, R.~H.~Brandenberger and Y.~K.~E.~Cheung,
  %``Big Bounce Genesis,''
  Phys.\ Rev.\ D {\bf 90} (2014) 12,  123535
  [arXiv:1403.5625 [gr-qc]].;\\
  %%CITATION = ARXIV:1403.5625;%%
  %7 citations counted in INSPIRE as of 24 Apr 2015
 Yi-Fu Cai, E. McDonough,
F. Duplessis, R. H. Brandenberger, JCAP 1310 (2013) 024 [arXiv:1305.5259];\\
Yi-Fu Cai, E. Wilson-Ewing, JCAP 1403 (2014) 026 [arXiv:1402.3009 ];\\
J. Haro, J. Amoros, JCAP 08(2014)025 [arXiv:1403.6396 ];\\
T.~Qiu and K.~C.~Yang,   %``Perturbations in Matter Bounce with Non-minimal Coupling,''
  JCAP {\bf 1011}, 012 (2010)
  [arXiv:1007.2571 [astro-ph.CO]].;\\
  %%CITATION = ARXIV:1007.2571;%%\bibitem{Qiu:2010vk}
  T.~Qiu,   %``Can the Big Bang Singularity be avoided by a single scalar field?,''
  Class.\ Quant.\ Grav.\  {\bf 27}, 215013 (2010)
  [arXiv:1007.2929 [hep-ph]].

\bibitem{bounce2} Jean-Luc Lehners, Class.Quant.Grav. {\bf 28} (2011) 204004 [arXiv:1106.0172]

\bibitem{bounce3} N. Arkani-Hamed, Hsin-Chia Cheng, M. A. Luty, S. Mukohyama, JHEP {\bf 0405} (2004) 074 [hep-th/0312099]; A. Nicolis, R. Rattazzi, E. Trincherini,
Phys.Rev. D {\bf 79} (2009) 064036 [arXiv:0811.2197];\\
 C. Deffayet , G.
Esposito-Farese (Paris, Inst. Astrophys.), A. Vikman, Phys.Rev. D {\bf 79}
(2009) 084003 [arXiv:0901.1314]; \\
J. Khoury, B. A. Ovrut, J. Stokes,
JHEP {\bf  1208} (2012) 015 [arXiv:1203.4562]



\bibitem{bounce4} M. Koehn, Jean-Luc Lehners, B. A. Ovrut, Phys.Rev. D {\bf 90} (2014) 025005 [arXiv:1310.7577]

\bibitem{bounce5} Yi-Fu Cai, D. A. Easson, R. Brandenberger, JCAP {\bf 1208} (2012) 020 [arXiv:1206.2382]; \\
%\bibitem{Odintsov:2015zua}
  S.~D.~Odintsov and V.~K.~Oikonomou,
  %``$\Lambda$CDM bounce cosmology without $\Lambda$CDM: the case of modified gravity,''
  Phys.\ Rev.\ D {\bf 91} (2015) 6,  064036
  [arXiv:1502.06125 [gr-qc]]
  %%CITATION = ARXIV:1502.06125;%%
'
\bibitem{quintombounce}  Y.~F.~Cai, T.~Qiu, Y.~S.~Piao, M.~Li and X.~Zhang,
  %``Bouncing universe with quintom matter,''
  JHEP {\bf 0710} (2007) 071
  [arXiv:0704.1090 [gr-qc]].
  %%CITATION = ARXIV:0704.1090;%%
  %154 citations counted in INSPIRE as of 22 Apr 2015



\bibitem{matterbounce}
R. H. Brandenberger, arXiv:1206.4196;\\
J.~Quintin, Y.~F.~Cai and R.~H.~Brandenberger,
%``Matter creation in a nonsingular bouncing cosmology,''
Phys.\ Rev.\ D {\bf 90} (2014) 6,  063507
[arXiv:1406.6049 [gr-qc]]; \\
%%CITATION = ARXIV:1406.6049;%%
%9 citations counted in INSPIRE as of 14 Feb 2015
%\cite{Cai:2011ci}
%\bibitem{Cai:2011ci}
Y.~F.~Cai, R.~Brandenberger and X.~Zhang,
%``Preheating a bouncing universe,''
Phys.\ Lett.\ B {\bf 703} (2011) 25
[arXiv:1105.4286 [hep-th]]; \\
%%CITATION = ARXIV:1105.4286;%%
%22 citations counted in INSPIRE as of 14 Feb 2015
%\cite{Cai:2011zx}
%\bibitem{Cai:2011zx}
Y.~F.~Cai, R.~Brandenberger and X.~Zhang,
%``The Matter Bounce Curvaton Scenario,''
JCAP {\bf 1103} (2011) 003
[arXiv:1101.0822 [hep-th]];\\
%%CITATION = ARXIV:1101.0822;%%
%43 citations counted in INSPIRE as of 14 Feb 2015
K.~Bamba, J.~de Haro and S.~D.~Odintsov,
%``Future Singularities and Teleparallelism in Loop Quantum Cosmology,''
JCAP {\bf 1302} (2013) 008
[arXiv:1211.2968 [gr-qc]];\\
J.~de Haro,
%``Does loop quantum cosmology replace the big rip singularity by a non-singular bounce?,''
JCAP {\bf 1211} (2012) 037
[arXiv:1207.3621 [gr-qc]].; \\
%\cite{Odintsov:2014gea}
%\bibitem{Odintsov:2014gea}
  S.~D.~Odintsov and V.~K.~Oikonomou,
  %``Matter Bounce Loop Quantum Cosmology from $F(R)$ Gravity,''
  Phys.\ Rev.\ D {\bf 90} (2014) 12,  124083
  [arXiv:1410.8183 [gr-qc]];\\
  J. Haro, Europhys. Lett. {\bf 107} (2014) 29001 [arXiv:1403.4529];\\
  %\cite{Oikonomou:2014yua}
Jaume de Haro, Yi-Fu Cai, arXiv:1502.03230;\\
K. Bamba, S. Nojiri, S. D. Odintsov, JCAP {\bf 0810} (2008) 045 [arXiv:0807.2575];\\
K.~Bamba, A.~N.~Makarenko, A.~N.~Myagky, S.~Nojiri and S.~D.~Odintsov,
  %``Bounce cosmology from $F(R)$ gravity and $F(R)$ bigravity,''
  JCAP {\bf 1401} (2014) 008
  [arXiv:1309.3748 [hep-th]]. ;\\
  C.~Barragan, G.~J.~Olmo and H.~Sanchis-Alepuz,
  %``Bouncing Cosmologies in Palatini f(R) Gravity,''
  Phys.\ Rev.\ D {\bf 80} (2009) 024016
  [arXiv:0907.0318 [gr-qc]].

\bibitem{superbounce2}
 S.~D.~Odintsov, V.~K.~Oikonomou and E.~N.~Saridakis,
  %``Superbounce and Loop Quantum Ekpyrotic Cosmologies from Modified Gravity: $F(R)$, $F(G)$ and $F(T)$ Theories,''
  arXiv:1501.06591 [gr-qc].
  %%CITATION = ARXIV:1501.06591;%%
  %3 citations counted in INSPIRE as of 22 Apr 2015

\bibitem{superbounce3}
  V.~K.~Oikonomou,
  %``Superbounce and Loop Quantum Cosmology Ekpyrosis from Modified Gravity,''
  arXiv:1412.4343 [gr-qc]
  %%CITATION = ARXIV:1412.4343;%%
  %3 citations counted in INSPIRE as of 31 Mar 2015


\bibitem{superbounce4}

 S.~D.~Odintsov and V.~K.~Oikonomou,
  %``Matter Bounce Loop Quantum Cosmology from $F(R)$ Gravity,''
  Phys.\ Rev.\ D {\bf 90} (2014) 12,  124083
  [arXiv:1410.8183 [gr-qc]].


\bibitem{ekpyr1} J. Khoury, B. A. Ovrut , N. Seiberg, P. J. Steinhardt, N. Turok, Phys.Rev. D {\bf 65} (2002) 086007 [hep-th/0108187];\\
J. K. Erickson, D. H. Wesley, P. J. Steinhardt, N. Turok, Phys.Rev. D {\bf 69} (2004) 063514 [hep-th/0312009]; \\
Yi-Fu Cai, Shih-Hung Chen, J. B. Dent, S. Dutta, E. N. Saridakis, Class. Quantum Grav. {\bf 28} (2011) 215011 [arXiv:1104.4349]

\bibitem{ekpyr2}  J. Khoury, B. A. Ovrut, P. J. Steinhardt, N. Turok, Phys.Rev. D {\bf 66} (2002) 046005 [hep-th/0109050]

\bibitem{ekpyr3} E. Wilson-Ewing, JCAP {\bf 1303} (2013) 026 [arXiv:1211.6269]



\bibitem{powerspectrum}
D.~Wands,
  Phys.\ Rev.\ D {\bf 60}, 023507 (1999)
  [arXiv:9809062]. \\
  F.~Finelli and R.~Brandenberger,
  Phys. Rev. {\bf D65}, 103522 (2002)
  [arXiv:0112249].\\
  R.~H.~Brandenberger and R.~Kahn,
  %``Cosmological Perturbations In Inflationary Universe Models,''
  Phys.\ Rev.\ D {\bf 29} (1984) 2172.\\
R.~H.~Brandenberger, R.~Kahn and W.~H.~Press,
  %``Cosmological Perturbations in the Early Universe,''
  Phys.\ Rev.\ D {\bf 28} (1983) 1809.\\
 V.~F.~Mukhanov, H.~A.~Feldman and R.~H.~Brandenberger,
  %``Theory of cosmological perturbations. Part 1. Classical perturbations. Part 2. Quantum theory of perturbations. Part 3. Extensions,''
  Phys.\ Rept.\  {\bf 215} (1992) 203.\\
J.~Haro and J.~Amoros,
  JCAP {\bf 1412}, no. 12, 031 (2014)
  [arXiv:1406.0369 ].\\
Y.~F.~Cai, D.~A.~Easson and R.~Brandenberger,
  JCAP {\bf 1208}, 020 (2012)
  [arXiv:1206.2382].\\
  J.~Haro,
  JCAP {\bf 1311}, 068 (2013)
  [Erratum-ibid.\  {\bf 1405}, E01 (2014)]
  [arXiv:1309.0352].\\
E.~Wilson-Ewing,
  JCAP {\bf 1303}, 026 (2013)
  [arXiv:1211.6269].\\
Y.~F.~Cai and E.~Wilson-Ewing,
  JCAP {\bf 1403}, 026 (2014)
  [arXiv:1402.3009].\\
Y.~F.~Cai,
  Sci.\ China Phys.\ Mech.\ Astron.\  {\bf 57}, 1414 (2014)
  [arXiv:1405.1369].\\
J.~de Haro and J.~Amoros,
  JCAP {\bf 1408}, 025 (2014)
  [arXiv:1403.6396].; \\
  R.~H.~Brandenberger, V.~F.~Mukhanov and A.~Sornborger,
  %``A Cosmological theory without singularities,''
  Phys.\ Rev.\ D {\bf 48} (1993) 1629
  [gr-qc/9303001].; \\
V.~F.~Mukhanov and R.~H.~Brandenberger,
  %``A Nonsingular universe,''
  Phys.\ Rev.\ Lett.\  {\bf 68} (1992) 1969.;\\
L.~Sebastiani and R.~Myrzakulov,
  %``F(R) gravity and inflation,''
  arXiv:1506.05330 [gr-qc].; \\
  S.~Myrzakul, R.~Myrzakulov and L.~Sebastiani,
  %``$f(\phi) R$-models for inflation,''
  arXiv:1509.07021 [gr-qc].


\bibitem{eho}
E.~Elizalde, J.~Haro and S.~D.~Odintsov, Phys. Rev. {\bf D91}, 063522 (2015)
  [arXiv:1411.3475].

\bibitem{planck}
 P.~A.~R.~Ade {\it et al.}  [Planck Collaboration],
  Astron.\ Astrophys.\  {\bf 571}, A16 (2014)
  [arXiv:1303.5076].\\
  P.~A.~R.~Ade {\it et al.}  [Planck Collaboration],
  Astron.\ Astrophys.\  {\bf 571}, A22 (2014)
  [arXiv:1303.5082].\\
  P.~A.~R.~Ade {\it et al.}  [Planck Collaboration],
  [arXiv:1502.01589].\\
  P.~A.~R.~Ade {\it et al.}  [Planck Collaboration],
  [arXiv:1502.02114].


\bibitem{oikofg}

 V.~K.~Oikonomou,
  %``Singular Bouncing Cosmology from Gauss-Bonnet Modified Gravity,''
  arXiv:1509.05827 [gr-qc]


\bibitem{dsy}
N. Deruelle, Y. Sendouda and A. Youssef, Phys. Rev. {\bf D80}, 084032 (2009)
  [arXiv:0906.4983].

  \bibitem{nojiri}
  S. Nojiri and S.D. Odintsov, Phys. Lett. {\bf B631}, 1 (2005) [arXiv:0508049].

  \bibitem{s09}
 P. Singh, Class. Quant. Grav. {\bf 26}, 125005 (2009)
  [arXiv:0901.2750].\\
  A.~Corichi and P.~Singh,
  %``Is loop quantization in cosmology unique?,''
  Phys.\ Rev.\ D {\bf 78}, 024034 (2008)
  [arXiv:0805.0136];
  \bibitem{h12}
J. de Haro, JCAP {\bf 11}, 037 (2012)
  [arXiv:1207.3621].

\bibitem{as11}
A. Ashtekar and P. Singh, Class. Quant. Grav. {\bf 28}, 213001 (2011)
  [arXiv:1108.0893].


\bibitem{aps06} A. Ashtekar, T. Pawlowski and P. Singh , Phys. Rev.  {\bf D74}, 084003 (2006).\\
A.~Ashtekar, T.~Pawlowski and P.~Singh,
  Phys.\ Rev. {\bf D73}, 124038 (2006)
  [arXiv:0604013].\\
 A. Ashtekar, M. Bojowald and J. Lewandowski, Adv. Theor. Math. {\bf 7}, 233 (2003).\\
T.~Thiemann,
  %``Anomaly - free formulation of nonperturbative, four-dimensional Lorentzian quantum gravity,''
  Phys.\ Lett.\ B {\bf 380}, 257 (1996)
  [arXiv:9606088]. \\
T.~Thiemann,
  Class.\ Quant.\ Grav.\  {\bf 15}, 839 (1998)
  [arXiv:9606089].\\
T.~Thiemann,
  Cambridge, UK: Cambridge Univ. Press. (2007)
  [arXiv:0110034].


\bibitem{bo08} M. Bojowald, Class. Quant. Grav. {\bf 26}, 075020 (2009) [arXiv:0811.4129].
\bibitem{he10} J. Haro and E. Elizalde, EPL {\bf 89}, 69001 (2010).
\bibitem{dmw09}  P. Dzierzak, P. Malkiewicz and W. Piechocki, Phys. Rev. {\bf D80}, 104001 (2009).

\bibitem{Friedmann}
P.~Singh, K.~Vandersloot and G.~V.~Vereshchagin,
  Phys.\ Rev. {\bf D74}, 043510 (2006)
  [arXiv:0606032].\\
P.~Singh,
  Phys.\ Rev. {\bf D73}, 063508 (2006)
  [arXiv:0603043]. \\
  C.~Rovelli and E.~Wilson-Ewing,
  Phys.\ Rev.\ D {\bf D90},  023538 (2014)
  [arXiv:1310.8654].





\bibitem{bojowald}
M. Bojowald and G.M. Hossain,
%Loop quantum gravity corrections to gravitational wave
%dispersion,
Phys. Rev. {\bf D77} 023508 (2008) [arXiv:0709.2365].\\
T. Cailleteau, J. Mielczarek, A. Barrau and J. Grain,
%Anomaly-free scalar perturbations with
%holonomy corrections in loop quantum cosmology,
Class. Quant. Grav. {\bf 29} 095010 (2012)
[arXiv:1111.3535].\\
T. Cailleteau, A. Barrau, J. Grain and F. Vidotto,
%Consistency of holonomy-corrected scalar,
%vector and tensor perturbations in Loop Quantum Cosmology,
Phys. Rev. {\bf D86}  087301
(2012) [arXiv:1206.6736].\\
T. Cailleteau, L. Linsefors and A. Barrau, Class.Quant.Grav. {\bf 31}  125011 (2014)
%Anomaly-free perturbations with inverse-volume and
%holonomy corrections in Loop Quantum Cosmology,
[arXiv:1307.5238].



















\bibitem{Fgg} Nojiri S., Odintsov S.D., Phys. Lett. {\bf B631}, 1 (2005) [arxiv:0508049].\\ Nojiri S., Odintsov
S.D., Sasaki M., Gauss-Bonnet dark energy. Phys. Rev. {\bf D71}, 123509 (2005) [arXiv:0504052].\\
Ratbay Myrzakulov, Diego Saez-Gomez, Anca Tureanu, Gen.Rel.Grav. {\bf 43}, 1671-1684 (2011).\\ E.
Elizalde, R. Myrzakulov, V. V. Obukhov and D. Saez-Gomez, Class. Quant. Grav. {\bf 27}, 095007 (2010)
[arXiv:1001.3636].\\ A.N. Makarenko, V.V. Obukhov, I.V. Kirnos, Astrophys.Space Sci. {\bf 343}, 481-488 (2013)  [arXiv:1201.4742].


\bibitem{sergeinojiri}
S.~Nojiri, S.~D.~Odintsov and S.~Tsujikawa,
  %``Properties of singularities in (phantom) dark energy universe,''
  Phys.\ Rev.\ D {\bf 71} (2005) 063004
  [hep-th/0501025].



\bibitem{sno2015}

J.~D.~Barrow and A.~A.~H.~Graham,
  %``Singular Inflation,''
  Phys.\ Rev.\ D {\bf 91}, no. 8, 083513 (2015)
  [arXiv:1501.04090 [gr-qc]].\\
S.~Nojiri, S.~D.~Odintsov and V.~K.~Oikonomou,
  %``Quantitative analysis of singular inflation with scalar-tensor and modified gravity,''
  Phys.\ Rev.\ D {\bf 91} 084059 (2015)
  [arXiv:1502.07005 [gr-qc]].\\
S.~Nojiri, S.~D.~Odintsov, V.~K.~Oikonomou and E.~N.~Saridakis,
  %``Singular cosmological evolution using canonical and phantom scalar fields,''
  arXiv:1503.08443 [gr-qc].













\end{thebibliography}
\end{document}